\documentclass[fleqn,usenatbib,onecolumn]{mnras}

\usepackage{newtxtext,newtxmath}

\usepackage[T1]{fontenc}

\usepackage{pdflscape}

\DeclareRobustCommand{\VAN}[3]{#2}
\let\VANthebibliography\thebibliography
\def\thebibliography{\DeclareRobustCommand{\VAN}[3]{##3}\VANthebibliography}

\usepackage{graphicx}	
\usepackage{amsmath}	

\usepackage{subcaption} 

\title[Fresnel Models for GW Effects on Pulsar Timing]{Fresnel Models for Gravitational Wave Effects on Pulsar Timing}

\author[McGrath \& Creighton]{
Casey McGrath,$^{1}$
and Jolien Creighton,$^{1}$\thanks{E-mail: jolien@uwm.edu}
\\
$^{1}$Center for Gravitation, Cosmology \& Astrophysics, Department of Physics, University of Wisconsin-Milwaukee, P.O. Box 413, Milwaukee, WI 53201, USA\\
}

\date{Accepted 2021 May 13. Received 2021 May 11; in original form 2020 November 18}

\pubyear{2021}

\begin{document}
\label{firstpage}
\pagerange{4531–4554}
\maketitle

\begin{abstract}
Merging supermassive black hole binaries produce low frequency gravitational waves, which pulsar timing experiments are searching for.  Much of the current theory is developed within the plane-wave formalism, and here we develop the more general Fresnel formalism.  We show that Fresnel corrections to gravitational wave timing residual models allow novel measurements to be made, such as direct measurements of the source distance from the timing residual phase and frequency, as well as direct measurements of chirp mass from a monochromatic source.  Probing the Fresnel corrections in these models will require future pulsar timing arrays with more distant pulsars across our Galaxy (ideally at the distance of the Magellanic Clouds), timed with precisions less than 100 ns, with distance uncertainties reduced to the order of the gravitational wavelength.  We find that sources with chirp mass of order $10^9 \ \mathrm{M}_\odot$ and orbital frequency $\omega_0 > 10$ nHz are good candidates for probing Fresnel corrections.  With these conditions met, the measured source distance uncertainty can be made less than 10 per cent of the distance to the source for sources out to $\sim 100$ Mpc, source sky localization can be reduced to sub-arcminute precision, and source volume localization can be made to less than $1 \ \text{Mpc}^3$ for sources out to $1$-Gpc distances.
\end{abstract}




    \section{Introduction}\label{sec:intro}

Supermassive black hole binaries (SMBHBs) at the centres of coalescing galaxies are expected to produce gravitational waves on the order $\mathcal{O}(1-100 \ \text{nHz})$.  Currently, pulsar timing collaborations such as the North American Nanohertz Observatory for Gravitational Waves (NANOGrav), the Parkes Pulsar Timing Array, and the European Pulsar Timing Array have been regularly collecting timing data on numerous pulsars for over a decade in order to search for the gravitational wave signal of these sources~\citep{PPTA_cw_2014,EPTA_cw_2016,NG_11yr_cw}.  Many pulsars have rotational periods that behave like extremely regular astrophysical clocks, but in the presence of a continuous gravitational wave, these observed periods will slowly oscillate in time.  The gravitational wave-induced period shifts in every rotation of the pulsar then stack over time, causing the arrival times of the pulsar clock signals to drift in and out of synchronization with a reference clock.  Therefore, by timing pulsars across our Galaxy in a pulsar timing array (PTA) and comparing the timing residual data with our theoretical models for these residuals, we can search for the presence of a gravitational wave and if found, infer the model parameters that describe the SMBHB source.  While the first detection may be the cumulative effect of gravitational waves from many sources across the sky as a stochastic background, here we focus on the possibility of detecting individual loud continuous wave signals coming from SMBHBs.

An important assumption that goes into deriving the gravitational wave timing residual model is the description of the shape of the gravitational wavefront.  Many studies such as~\citet{CC_main_paper} use models that assume that the PTA receives plane-waves coming from distant sources, with the notable exception of~\citet{DF_main_paper}, which considered a curved wavefront.  A second important assumption that is built into these models is the evolution of the gravitational wave in time.  Models can either treat the SMBHB as a purely monochromatic gravitational wave source~\citep{GWastro_Lee2011} or include the frequency evolution of the gravitational waves over the thousands to tens of thousands of years it takes the radio waves to travel from the pulsar to the Earth (\citeauthor{CC_main_paper}).

\begin{figure}
    \centering
      \begin{subfigure}[t]{0.48\linewidth}
      \centering
        \includegraphics[width=1\linewidth]{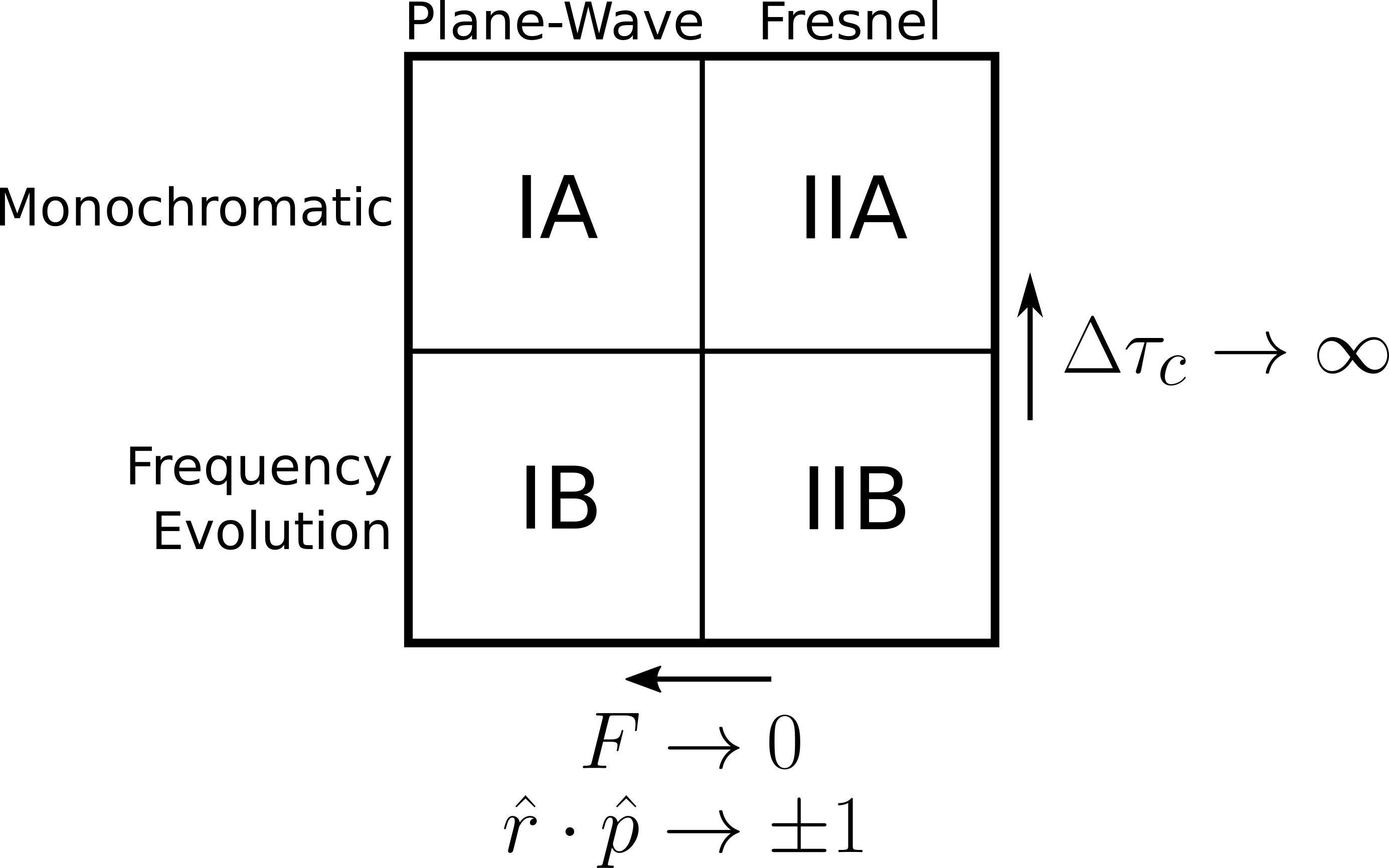}
      \end{subfigure}
      \hfill
      \begin{subfigure}[t]{0.4\linewidth}
      \centering
      \includegraphics[width=1\linewidth]{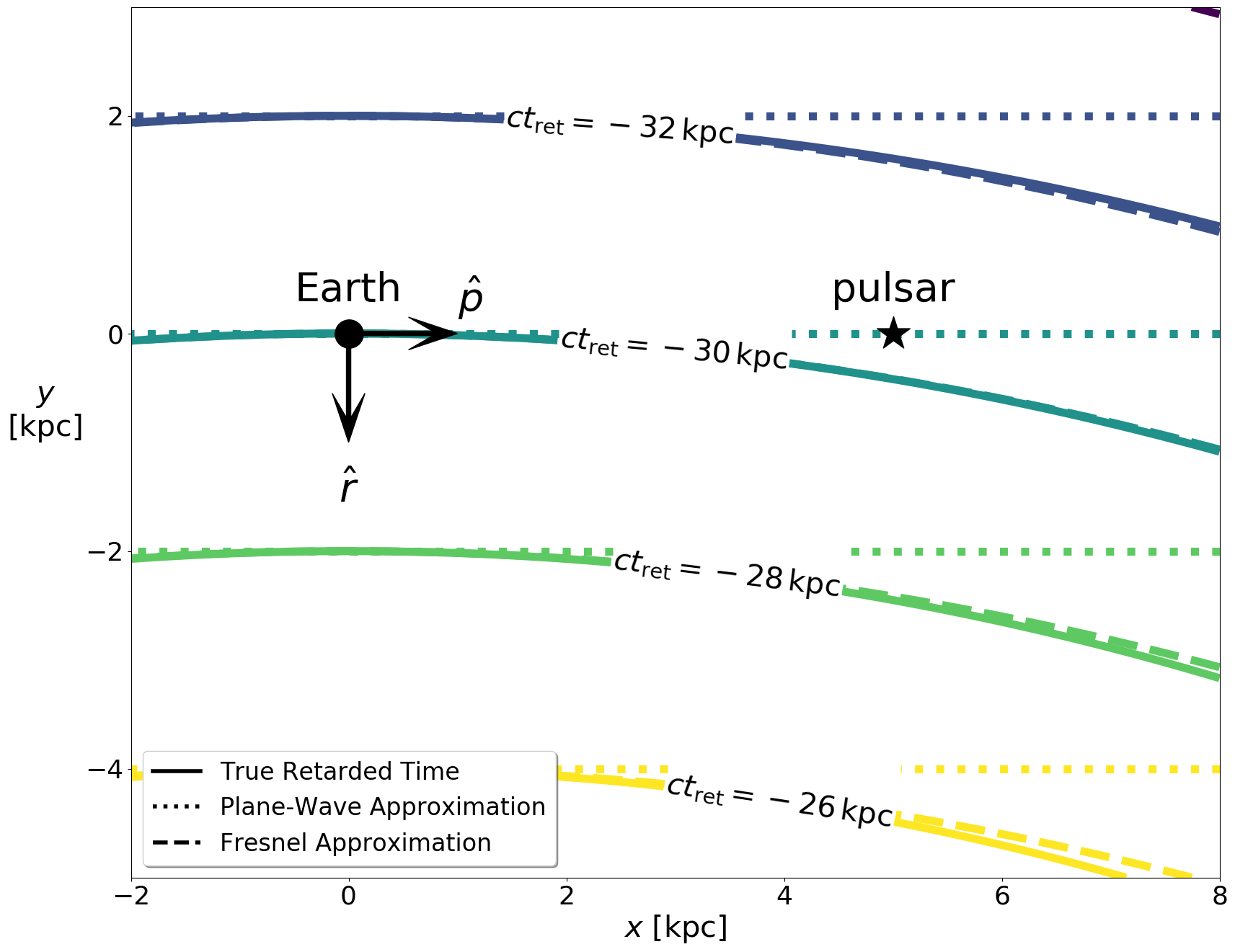}
      \end{subfigure}
    \caption{\textbf{Left:} Our classification of the gravitational wave regimes.  They increase in generality from left to right and top to bottom.  The models reduce from frequency evolution to monochromatism in the large coalescence time limit ($\Delta \tau_c \rightarrow \infty$) and from Fresnel to plane-wave in either the small Fresnel number limit ($F \rightarrow 0$) or the natural plane-wave limit ($\hat{r}\cdot\hat{p} \rightarrow \pm 1$).  The plane-wave regime has been well studied in pulsar timing literature, and in this paper, we add a new `Fresnel' regime.   \textbf{Right:} An example contour diagram of the retarded time in the source-Earth-pulsar plane at fixed time $t=0$.  The approximations indicate up to what term is included in the expansion in equation~\ref{eqn:retarded time}.  Since the lines show constant $t_\mathrm{ret}$, this traces the curvature of the wavefront.  For illustrative purposes that exaggerate and show the differences in the approximation regimes, we set the pulsar at a distance $L=5$ kpc and the source at $R=30$ kpc from the Earth.  A source with frequency $\omega_0=1$ nHz along with this pulsar would have a Fresnel number $F=27.3$ in this example.  As we can see here, the curvature of the wavefront means that the wave arrives at the pulsar's location \textit{later} than what is predicted by the plane-wave approximation.}
\label{fig: 4 regimes / tret contours}
\end{figure}

In this paper, we organize and classify four regimes of interest that we show in the left-hand panel of Figure~\ref{fig: 4 regimes / tret contours}, which increase in generality left to right and top to bottom and govern the fundamental assumptions built into the timing residual models.  Of these four regimes, the plane-wave models are well established in previous literature.  We add a new regime, which we label `Fresnel,' as we will show that it becomes important for significant Fresnel numbers describing the curvature of wavefronts.  To this, we derive a formula for the timing residuals in the Fresnel monochromatic regime and use it to propose a physically motivated conjecture as to what the most general gravitational wave residual model should be, the Fresnel frequency evolution regime.

The inspiration of this work comes primarily from~\citet{DF_main_paper} and~\citet{CC_main_paper}.  Our primary goal is to investigate how well we can measure the source distance parameter purely from the Fresnel corrections due to the curvature of the gravitational wave in our models, like in~\citeauthor{DF_main_paper}.  We adopt a Fisher matrix approach as in~\citeauthor{CC_main_paper} to perform this analysis, and we also include the pulsar distances as free parameters in our studies following their approach.  

For this first study, we do emphasize that in order to focus on the measurement of the source distance and to try and determine what experimental PTA requirements we will need for this measurement to be feasible, we adopt a Euclidean cosmology and assume a static universe.  This means that there is only one type of distance measurement of the physical distance between the Earth and the source in the work presented here.  In a more generalized cosmology, the source distance parameter that is searched for in timing residual models when working in the plane-wave regime is the luminosity distance~\citep{PPTA_cw_2014,EPTA_cw_2016,NG_11yr_cw}, which comes from the amplitude of the timing residual.  However, the curvature corrections from the Fresnel regime introduce a second distance parameter into the timing residual models, the comoving distance.  In this paper, we do not attempt to measure the source distance from the amplitude of the timing residual, only from the wavefront curvature corrections.  Therefore, the distance we recover would represent the comoving distance (not the luminosity distance) in a more general cosmological framework.  The idea of measuring both distances in a generalized cosmological model was recently explored in~\citet{DOrazio_main_paper} as a means of measuring the Hubble constant using PTA experiments, and it will also be the follow-up study to this work, using the formalism and results that we establish in this paper.

It is expected that measuring the comoving distance from the Fresnel corrections will be more difficult than measuring the luminosity distance from the amplitude, which will therefore limit the ability to measure cosmological parameters solely with gravitational waves.  In this study, we show that the source sky and volume localization from measurements of the gravitational wave source parameters are greatly improved by highly accurate measurements of the pulsar distances in our PTA.  Localizing the source could be invaluable in the identification of an optical counterpart.  If a counterpart is found in a galaxy of known redshift, then gravitational waves would provide a standard siren that would determine the distance of the source and hence a measurement of the Hubble constant~\citep{Schutz_1986,Holz_2005}.

The paper is outlined as follows.  In Section~\ref{sec:timing residual theory}, we present a new notational description of gravitational wave timing residuals in pulsars, which we think helps to clarify and emphasize the underlying physics and assumptions that go into building the timing residual models.  Then, we develop the actual residual models in Section~\ref{sec:4 model regimes}, derive the Fresnel monochromatic regime formula, and motivate the full Fresnel frequency evolution regime.  In Section~\ref{sec:fisher matrix analysis}, we explain our Fisher matrix method for parameter estimation, and in Section~\ref{sec:L-wrapping problem}, we provide a fresh look at the pulsar distance wrapping problem and its importance in parameter estimation.  The results of our Fisher analysis are then given in Section~\ref{sec:results}, which focus on the new insights that the Fresnel regime offers to pulsar timing.  Our final conclusions and future directions of this work are then given in Section~\ref{sec: conclusions}.

    \section{Gravitational Wave Timing Residual Theory}\label{sec:timing residual theory}

For this study, we assume a flat and static universe.  Additional assumptions are listed in Appendix~\ref{app: assumptions}.  There are eight source parameters for a circular binary system in the zeroth-order Newtonian regime $\{R, \theta, \phi, \iota, \psi, \theta_0, \mathcal{M}, \omega_0 \}$.  These are the Earth-source distance (recall that we are assuming flat static space), sky angles, orientation Euler angles (inclination and polarization), initial phase (which can also be interpreted as the third orientation Euler angle), system chirp mass $\mathcal{M}\equiv \left(m_1 m_2\right)^{3/5} / \left(m_1 + m_2\right)^{1/5}$ (a combination of the individual black hole masses), and orbital frequency parameter.  The pulsar parameters are $\{L, \theta_p, \phi_p \}$, which are the Earth-pulsar distance, and pulsar sky angles.

We choose to use the following notation and conventions in this work.  With the Earth at the centre of the coordinate system, the source frame orientation vectors and the pulsar vector are defined by (see also the right-hand panel of Figure~\ref{fig: 4 regimes / tret contours}):
\begin{equation}
    \begin{cases}
        \begin{tabular}{c l c r l}
            $\hat{r}$ &$\equiv \big[\sin(\theta)\cos(\phi),$ &$\sin(\theta)\sin(\phi),$ &$\cos(\theta) \big]$ \ , & (Earth to gravitational wave source unit vector) \\
            $\hat{\theta}$ &$\equiv \big[\cos(\theta)\cos(\phi),$ &$\cos(\theta)\sin(\phi),$ &$\sin(\theta) \big]$ \ , & (transverse plane basis vector) \\
            $\hat{\phi}$ &$\equiv \big[-\sin(\phi),$ &$\cos(\phi),$ &$0 \big]$ \ , & (transverse plane basis vector) \\
            $\hat{p}$ &$\equiv \big[\sin(\theta_p)\cos(\phi_p),$ &$\sin(\theta_p)\sin(\phi_p),$ & $\cos(\theta_p) \big]$ \ , & (Earth to pulsar unit vector)
        \end{tabular}
        \end{cases} \label{eqn:source/pulsar basis vectors} 
\end{equation}
which we use to define the following generalized polarization tensors:
\begin{align}
    \begin{cases}
        e^{\hat{r}+}_{ij} &\equiv \hat{\theta}_i\hat{\theta}_j - \hat{\phi}_i\hat{\phi}_j, \\
        e^{\hat{r}\times}_{ij} &\equiv \hat{\phi}_i\hat{\theta}_j + \hat{\theta}_i\hat{\phi}_j, \\
        E^{\hat{r}+}_{ij} &\equiv \frac{1}{2}\left(1 + \cos^2(\iota)\right) \left[\cos(2\psi)e^{\hat{r}+}_{ij} + \sin(2\psi)e^{\hat{r}\times}_{ij} \right], \\
        E^{\hat{r}\times}_{ij} &\equiv \cos(\iota) \left[-\sin(2\psi)e^{\hat{r}+}_{ij} + \cos(2\psi)e^{\hat{r}\times}_{ij} \right].
    \end{cases} \label{eqn: generalized polarization tensors}
\end{align}
Writing the polarization tensors in this way groups all of the geometrical orientation and location angles into the definition of the tensors and keeps them from mixing the plus and cross-metric perturbations, which we find to be conceptually and mathematically convenient.  Another important quantity we will define here are the following antenna patterns:
\begin{align}
    \begin{cases}
        \begin{tabular}{l l l}
            $f^{+}$ &$\equiv \frac{\hat{p}^i\hat{p}^j e^{\hat{r}+}_{ij}}{\left(1-\hat{r}\cdot\hat{p}\right)}$ &$= \frac{\hat{p}^i\hat{p}^j\hat{\theta}_i\hat{\theta}_j - \hat{p}^i\hat{p}^j\hat{\phi}_i\hat{\phi}_j}{\left(1-\hat{r}\cdot\hat{p}\right)} \quad = \frac{\left(\hat{p}\cdot\hat{\theta}\right)^2 - \left(\hat{p}\cdot\hat{\phi}\right)^2}{\left(1-\hat{r}\cdot\hat{p}\right)}$ , \\[3pt]
            $f^{\times}$ &$\equiv \frac{\hat{p}^i\hat{p}^j e^{\hat{r}\times}_{ij}}{\left(1-\hat{r}\cdot\hat{p}\right)}$ &$= \frac{\hat{p}^i\hat{p}^j\hat{\phi}_i\hat{\theta}_j + \hat{p}^i\hat{p}^j\hat{\theta}_i\hat{\phi}_j}{\left(1-\hat{r}\cdot\hat{p}\right)} \quad = \frac{2\left(\hat{p}\cdot\hat{\theta}\right)\left(\hat{p}\cdot\hat{\phi}\right)}{\left(1-\hat{r}\cdot\hat{p}\right)}$ , \\[3pt]
            $F^+$ &$\equiv \frac{\hat{p}^i\hat{p}^j E^{\hat{r}+}_{ij}}{\left(1-\hat{r}\cdot\hat{p}\right)}$ &$= \frac{1}{2}\left(1+\cos^2(\iota)\right) \left[ \cos(2\psi) f^+ + \sin(2\psi) f^\times \right]$ , \\[3pt]
            $F^\times$ &$\equiv \frac{\hat{p}^i\hat{p}^j E^{\hat{r}\times}_{ij}}{\left(1-\hat{r}\cdot\hat{p}\right)}$ &$= \cos(\iota) \left[ -\sin(2\psi) f^+ + \cos(2\psi) f^\times \right]$ .
        \end{tabular}
    \end{cases} \label{eqn: antenna patterns}
\end{align}
These will be used when writing the residual models in Section~\ref{sec:4 model regimes} and help to decide the detector sensitivity to the gravitational wave based on the Earth-pulsar-source geometrical alignment.  Note that in practice, the pulsar's pulse arrival times are referenced to the Solar System barycentre, so the barycentre would be the origin for the natural coordinate system.

Now the metric perturbation produced by a circular binary system can be written as~\citep{maggiore_2008,creighton_anderson_2011}:
\begin{align}
    h^{TT\hat{r}}_{ij} &= E^{\hat{r}+}_{ij}h_{+} + E^{\hat{r}\times}_{ij}h_{\times} = E^{\hat{r}\textsc{A}}_{ij}h_{\textsc{A}}  \qquad \mathrm{for}\quad \textsc{A} \in [+, \times],  \label{eqn: general metric perturbation} \\
    &\mathrm{where}\quad \begin{cases}
                h_{+}(t) &\equiv -h(t) \cos\big(2\Theta(t)\big), \\
                h_{\times}(t) &\equiv -h(t) \sin\big(2\Theta(t)\big), \\
                h(t) &\equiv \frac{4(G\mathcal{M})^{5/3}}{c^4 R}\omega(t)^{2/3} ,
            \end{cases} \label{eqn: h+x(t) & h(t)}
\end{align}
with the angular phase and frequency functions $\Theta(t)$ and $\omega(t)$ defined in Sections~\ref{subsec: mono vs freq evo}.

A more detailed description of the connection between the gravitational wave metric perturbation, induced fractional shift of a pulsar's period $T$, and its timing residual is offered in Appendix~\ref{app: deriving the timing residual}.  The gravitational wave-induced fractional shift of the pulsar's period $T$ is:
\begin{equation}
    \frac{\Delta T}{T}(t_\mathrm{obs}) \approx \frac{1}{2}\hat{p}^i\hat{p}^j E^{\hat{r}\textsc{A}}_{ij} \int\limits^{t_\mathrm{obs}}_{t_\mathrm{obs} - \frac{L}{c}} \frac{\partial h_{\textsc{A}}\big(t_\mathrm{ret}(t,\vec{x})\big)}{\partial t}\Bigg\rvert_{\vec{x}=\vec{x}_{0}(t)} dt ,
\label{eqn: Delta T / T}
\end{equation}
where $t_\mathrm{obs}$ is the time a pulsar's photon is observed arriving at Earth, $t_\mathrm{ret}$ is the retarded time of the gravitational wave (equation~\ref{eqn:retarded time}) , and $\vec{x}_{0}(t)$ is the spatial path of the photon between the pulsar and the Earth.

Finally, the gravitational wave-induced timing residual is the integrated fractional period shift due to the gravitational wave over the observation time.  Conceptually, this is the difference between the observed and expected time of arrival of a pulsar's pulse~\citep{creighton_anderson_2011, maggiore_2018}:
\begin{equation}
    \mathrm{Res}(t) = \int\frac{\Delta T}{T}\left(t_\mathrm{obs}\right) dt_\mathrm{obs} = \int\frac{T_\mathrm{obs} (t_\mathrm{obs}) - T}{T} dt_\mathrm{obs}  = \ \mathrm{Obs}(t) - \mathrm{Exp}(t) , \label{eqn:timing residual} \\
\end{equation}


    \subsection{Regimes: Monochromatism (A) versus frequency evolution (B)}\label{subsec: mono vs freq evo}
    
    In the simplest model of the gravitational waves emitted by a binary system, one neglects the energy loss by emission of gravitational radiation~\citep{maggiore_2008}.  Under this assumption, the binary's energy will remain conserved, the orbit will not coalesce (it has `infinite' coalescence time), and hence the orbital period/frequency will remain constant.  This assumption means that the gravitational waves emitted by the binary system will be monochromatic (their amplitudes and frequencies remain constant for all time).
    
    For the monochromatic gravitational wave model, the orbital frequency and phase are given by:
    \begin{align}
        \begin{cases}
            \omega(t) &= \omega_0 , \\ 
            \Theta(t) &= \theta_0 + \omega_0 (t - t_0) , 
        \end{cases}
        \label{eqn: monochrome model}
    \end{align}
    where $t_0$ here (and below) denotes the fiducial time for the model.  In this work, we will choose the fiducial time to be $t_0=-\frac{R}{c}$ (note that $t=0$ would correspond to the present-day start of our experiment on Earth).
    
    More realistically, however, energy in the binary system is lost due to the emission of gravitational wave radiation.  This means that over time, the binary will coalesce.  As the binary coalesces, the orbital frequency will increase, meaning that the gravitational wave frequency will also increase.  This evolution towards higher frequencies is called `frequency chirping.'  Due to the thousands of years it takes radio light to travel from the pulsar to the Earth, it is possible that the gravitational waves emitted from our observed source could evolve appreciably during this time.
    
    The orbital frequency and phase of the frequency evolving gravitational wave model are given by~\citep{maggiore_2008, creighton_anderson_2011, Arzoumanian_2014}:
    \begin{align}
        \begin{cases}
            \omega(t) &= \omega_0\left[1 - \frac{t-t_0}{\Delta\tau_c} \right]^{-3/8} , \\
            \Theta(t) &= \theta_0 + \theta_c\left[ 1 - \left(\frac{\omega(t)}{\omega_0}\right)^{-5/3} \right] . 
        \end{cases}
    \label{eqn: frequency evolution model}
    \end{align}
    Because a frequency-evolving source will eventually coalesce, the model is governed by the physically significant quantities $\Delta \tau_c$, which is the `coalescence time,' and $\theta_c$, which is the `coalescence angle' (the total angle swept out before the system coalesces).  Both of these quantities are measured from the chosen fiducial time and are given by:
    \begin{align}
        \begin{cases}
            \begin{tabular}{c l l}
                $\hspace{-0.2cm} \Delta\tau_c$ &$\equiv \frac{5}{256}\left(\frac{c^3}{G\mathcal{M}}\right)^{5/3} \frac{1}{\omega_0^{8/3}}$ &$\approx (430 \text{ Myr}) \left(\frac{10^9 M_\odot}{\mathcal{M}}\right)^{5/3} \left(\frac{1 \text{ nHz}}{\omega_0}\right)^{8/3}$ , \\
                $\hspace{-0.2cm} \theta_c$ &$\equiv \frac{8}{5}\Delta\tau_c \omega_0 \ \ =\ \  \frac{1}{32}\left(\frac{c^3}{G\mathcal{M}\omega_0}\right)^{5/3}$ &$\approx (3.5\times10^{6} \text{ cycles}) \left(\frac{10^9 M_\odot}{\mathcal{M}}\right)^{5/3} \left(\frac{1 \text{ nHz}}{\omega_0}\right)^{5/3}$ .
            \end{tabular}
        \end{cases}
    \label{eqn: coalescence time/angle}
    \end{align}
    Note that in the `large coalescence time' limit ($\Delta\tau_c \rightarrow \infty$), the frequency evolution model (equation~\ref{eqn: frequency evolution model}) reduces to the monochromatic model (equation~\ref{eqn: monochrome model}) as we would expect.  In practice, the monochromatic limit for pulsar timing experiments merely requires that the gravitational wave frequency stays nearly constant as pulsar's radiation travels between the pulsar and the Earth, so that $\Delta \tau_c \gg \left(1-\hat{r}\cdot\hat{p}\right)\frac{L}{c}$ (see equation~\ref{eqn: Res(t) pw freq evo phase E and P}).


    \subsection{Regimes: Plane-wave (I) vs. Fresnel (II)}\label{subsec: PW vs Fresnel}

    The wavefront of a travelling wave is traced by the retarded time, as shown in the right-hand panel of Figure~\ref{fig: 4 regimes / tret contours}.  For a static flat universe, the retarded time is:
    \begin{equation}
        t_\mathrm{ret} \quad = \quad t - \frac{|\vec{x} - \vec{x}^{'}|}{c} \quad = \quad t \quad - \underbrace{\frac{|\vec{x}^{'}|}{c}}_{\mathrm{``Far} \ \mathrm{Field"}} \quad + \quad \underbrace{\left(\hat{x}^{'}\cdot\hat{x}\right)\frac{|\vec{x}|}{c}}_{\mathrm{``Plane-Wave"}} \quad - \quad \underbrace{\frac{1}{2}\left(1-\left(\hat{x}^{'}\cdot\hat{x}\right)^2\right)\frac{|\vec{x}|}{c}\frac{|\vec{x}|}{|\vec{x}^{'}|}}_{\mathrm{``Fresnel"}} \quad + \quad \ldots 
    \label{eqn:retarded time}
    \end{equation}
    where $\vec{x}^{'}$ is the wave source's position, $\vec{x}$ is the field point of interest, and Earth is the origin.  In this Taylor expansion for $|\vec{x}^{'}| \gg |\vec{x}|$, we label the first three terms that are common in wave physics and optics.  This gives us a way of categorizing different wave solution approximation regimes, which will ultimately be the crux of our work in this paper.
    
    The Fresnel term gives the first-order curvature of the physical wavefront.  Conceptually, it introduces an additional time delay to the arrival time of the wavefront predicted by the plane-wave approximation (see the right-hand panel of Figure~\ref{fig: 4 regimes / tret contours}).  We want to understand when this curvature term/time delay will become significant in our timing residual model.  For an order-of-magnitude estimate, consider a monochromatic wave.  With our source at $\vec{x}^{'} =  R\hat{r}$ and the pulsar at $\vec{x} = L\hat{p}$, the timing residual solution (see Section~\ref{sec:4 model regimes}) will contain terms that behave like $\sin\left(2\omega_0 t_\mathrm{ret}\right)$.  These terms cycle on the interval $0\rightarrow 2\pi$, so the Fresnel term in equation~\ref{eqn:retarded time} will become significant when it is on the order of a cycle, that is roughly when $2\pi \sim \omega_0\left(1-\left(\hat{r}\cdot\hat{p}\right)^2\right)\left(\frac{L}{c}\right)\left(\frac{L}{R}\right) \equiv \left(1-\left(\hat{r}\cdot\hat{p}\right)^2\right)\pi F$, where:
    \begin{align}
        F &\equiv \frac{L^2}{\lambda_\mathrm{gw} R} \approx 0.0003 \left(\frac{\omega_0}{1\text{ nHz}}\right) \left(\frac{L}{1\text{ kpc}}\right)^2 \left(\frac{100\text{ Mpc}}{R}\right) , \label{eqn: Fresnel number} \\
        &\mathrm{and}\quad \lambda_\mathrm{gw} \equiv \frac{\pi c}{\omega_0} \approx (30.5 \text{ pc}) \left(\frac{1\text{ nHz}}{\omega_0}\right), \label{eqn: gw wavelength}
    \end{align}
    which is the standard definition of the Fresnel number.  If we ignore the geometric terms here, then when $F \sim \mathcal{O}(1)$ the Fresnel term in equation~\ref{eqn:retarded time} will contribute significantly, suggesting that the simple plane-wave approximation will be less accurate and that the Fresnel approximation will be required.  \citet{DF_main_paper} reached this same conclusion in their work.  Even when there exists significant frequency evolution in the timing residual model, the definition of the Fresnel number still offers us a convenient proxy of the importance of wavefront curvature within the model itself.
    
    There are two important limits to keep in mind here.  The `small Fresnel number' limit ($F \rightarrow 0$) and the `natural plane-wave limit' ($\hat{r}\cdot\hat{p} \rightarrow \pm 1$).  In either limit, we expect to recover the plane-wave result.  In the natural plane-wave limit, the source, Earth, and pulsar become perfectly aligned ($\hat{r}\cdot\hat{p} = 1$) or anti-aligned ($\hat{r}\cdot\hat{p} = -1$), which means that no amount of curvature in the wavefront would affect the timing any differently than a plane-wave.

    \section{Four Model Regimes}\label{sec:4 model regimes}

Solving equations~\ref{eqn: Delta T / T} and~\ref{eqn:timing residual} under the various assumptions and approximations in Sections~\ref{subsec: mono vs freq evo} and~\ref{subsec: PW vs Fresnel} gives four different gravitational wave timing residual models, which we compare here.  Some derivation detail and references are provided in Appendix~\ref{app: deriving the timing residual}.

Numerous sources use alternative notations and conventions when expressing these models.  For example, see~\citet{GWastro_Lee2011}, \citet{NG_11yr_cw}, \citet{PPTA_cw_2014}, \citet{EPTA_cw_2016}, \citet{Arzoumanian_2014}, and~\citet{Ellis_2013} for alternative ways of writing the residual model in the plane-wave frequency evolution regime.  One goal of this paper is simply to present the familiar results in a new light to make them more transparent and streamlined.  Furthermore, direct comparison of the models between the regimes in this notation helps to provide insight into the physical difference between each model and what is actually changing mathematically between them.


    \subsection{Plane-wave, monochromatic (IA)}\label{subsec: plane-wave, monochromatic}
    
    In the plane-wave monochromatic gravitational wave regime, the timing residual can be expressed compactly as:
    \begin{align}
        \mathrm{Res}(t) &= \frac{F^\textsc{A}}{4\omega_0} \bigg[ h_\textsc{A}\Big(\Theta_E-\frac{\pi}{4}\Big) - h_\textsc{A}\Big(\Theta_P-\frac{\pi}{4}\Big) \bigg]  \qquad \mathrm{for}\quad \textsc{A} \in [+, \times], \label{eqn: Res(t) pw mono} \\[4pt]
        &\underset{\left(t_0 \ = \ -\frac{R}{c}\right)}{\mathrm{where}}\quad \begin{cases}
            \Theta_E \equiv \Theta\Big(t-\frac{R}{c}\Big) &\equiv \theta_0 + \omega_0 t , \\
            \Theta_P \equiv \Theta\Big(t-\frac{R}{c}-\left(1-\hat{r}\cdot\hat{p}\right)\frac{L}{c}\Big) &\equiv \theta_0 + \omega_0\Big(t-\left(1-\hat{r}\cdot\hat{p}\right)\frac{L}{c}\Big) ,
            \end{cases}\label{eqn: Res(t) pw mono phase E and P} \\
        &\qquad\qquad \ \ \ \begin{cases}
            h_{+} \equiv -h_0 \cos\big(2\Theta\big), \\
            h_{\times} \equiv -h_0 \sin\big(2\Theta\big), \\
            h_0 \equiv \frac{4(G\mathcal{M})^{5/3}}{c^4 R}\omega_0^{2/3} ,
        \end{cases} \label{eqn: h0+x and h0}
    \end{align}
    along with equations~\ref{eqn:source/pulsar basis vectors}-\ref{eqn: antenna patterns}.  The term in brackets is the difference in the metric perturbation at the Earth and pulsar locations when the pulsar's radiation first left the pulsar and when it finally arrived at the Earth (denoted with `E' and `P' subscripts here and below).  The residual in the plane-wave regime depends only on the endpoints of the photon's motion, an `Earth term' and a `pulsar term.'  Note that in this regime, the source chirp mass $\mathcal{M}$ and distance $R$ parameters are fully degenerate in the residual.  Each parameter appears only in the combination $\frac{\mathcal{M}^{5/3}}{R}$ with the other (see equation~\ref{eqn: h0+x and h0}); therefore, we would not be able to measure these parameters independently through actual observations (only this specific combination of the two).
    
    An alternative but equivalent way of writing equation~\ref{eqn: Res(t) pw mono} is:
    \begin{align}
        \mathrm{Res}(t) &= A_\mathrm{c(IA)} \ F^\textsc{A} \ s_{E,\textsc{A}}  \qquad \mathrm{for}\quad \textsc{A} \in [+, \times], \label{eqn: Res(t) pw mono - alternative expression} \\[4pt]
        &\mathrm{where}\quad \begin{cases}
            A_\mathrm{c(IA)} \equiv \frac{h_\mathrm{c(IA)}}{4\omega_0} , \\
            h_\mathrm{c(IA)} \equiv 2 h_0 \sin\left(\omega_0\left(1-\hat{r}\cdot\hat{p}\right)\frac{L}{c}\right) ,
            \end{cases}\label{eqn: Res(t) pw mono characteristic strain and amplitude} \\
        &\qquad\quad \ \ \begin{cases}    
            s_{E,+} \equiv \sin\left(2\Theta_E - \frac{\pi}{2}-\omega_0\left(1-\hat{r}\cdot\hat{p}\right)\frac{L}{c}\right) , \\
            s_{E,\times} \equiv -\cos\left(2\Theta_E - \frac{\pi}{2}-\omega_0\left(1-\hat{r}\cdot\hat{p}\right)\frac{L}{c}\right) .
            \end{cases}
    \end{align}    
    The benefit of this notation is that it allows us to define in equation~\ref{eqn: Res(t) pw mono characteristic strain and amplitude} a `characteristic strain' $h_\mathrm{c(IA)}$ and `characteristic timing residual amplitude' $A_\mathrm{c(IA)}$ for the IA regime.  These characteristic terms will be useful when comparing this model to the Fresnel monochromatic model IIA in Section~\ref{subsubsec: fresnel, mono (Heuristic model)}.
    
     Note that the plane-wave monochromatic model converges to zero in \textit{both} of the natural plane-wave limits (when $\hat{r}\cdot\hat{p} = \pm 1$).  As the system becomes anti-aligned, the antenna patterns $F^\textsc{A}\rightarrow 0$.  As the system becomes aligned, the pulsar phase $\Theta_P \rightarrow \Theta_E$ in equation~\ref{eqn: Res(t) pw mono phase E and P}, and the difference in the metric perturbations at the Earth and the pulsar goes to zero fast enough to kill the entire timing residual (conceptually, the photons `surf' the gravitational waves in this second case).  In equation~\ref{eqn: Res(t) pw mono - alternative expression}, this is reflected in the characteristic strain and amplitudes themselves, which both go to zero in the alignment case.


    \subsection{Plane-wave, frequency evolution (IB)}\label{subsec: plane-wave, freq evolution}
    
    In the plane-wave frequency evolution gravitational wave regime, the timing residual can be expressed compactly as:
    \begin{align}
        \mathrm{Res}(t) &= \frac{F^\textsc{A}}{4} \left[ \frac{h_\textsc{A}\Big(\omega_{0E},\Theta_E-\frac{\pi}{4}\Big)}{\omega_{0E}} - \frac{h_\textsc{A}\Big(\omega_{0P},\Theta_P-\frac{\pi}{4}\Big)}{\omega_{0P}} \right] \qquad \mathrm{for}\quad \textsc{A} \in [+, \times], \label{eqn: Res(t) pw freq evo} \\[4pt]
        &\underset{\left(t_0 \ = \ -\frac{R}{c}\right)}{\mathrm{where}}\quad \begin{cases}
            \omega_{0E} &\equiv \omega_0 , \\
            \omega_{0P} &\equiv \omega_0\left[1 + \frac{\left(1-\hat{r}\cdot\hat{p}\right)\frac{L}{c}}{\Delta\tau_c}  \right]^{-3/8} , \\
            \theta_{0E} &= \theta_0 , \\
            \theta_{0P} &= \theta_0 + \theta_c\left(1-\left[1+ \frac{\left(1-\hat{r}\cdot\hat{p}\right)\frac{L}{c}}{\Delta\tau_c}\right]^{5/8}\right) , \\
            \Theta_E & = \theta_{0E} + \omega_{0E}t , \\
            \Theta_P & = \theta_{0P} + \omega_{0P}t ,
            \end{cases}\label{eqn: Res(t) pw freq evo phase E and P}
    \end{align}
    along with equations~\ref{eqn:source/pulsar basis vectors}-\ref{eqn: antenna patterns}, \ref{eqn: h+x(t) & h(t)}, and~\ref{eqn: coalescence time/angle}.  Like in the monochromatic case, this model depends only on the endpoints of the photon's motion.  However, in the case that frequency evolution is non-negligible, there is a slight difference in the frequency of the gravitational wave at the pulsar when the pulsar's radiation first left and the frequency of the gravitational wave at the Earth when the pulsar's radiation finally arrived.  Because the Earth and pulsar terms combine, in the cases where frequency evolution is non-negligible, we expect this to produce `beats' in the residual data that we observe.
    
    Equation~\ref{eqn: Res(t) pw mono} is recovered from equation~\ref{eqn: Res(t) pw freq evo} in the large coalescence time limit.  While this regime allows for frequency evolution over the thousands of years between emission and observation of the pulsar's pulses, it still assumes that frequency evolution is negligible over the experiment's observation time-scale.  Notice that in this regime the degeneracy of the source chirp mass $\mathcal{M}$ and distance $R$ parameters is now broken since $\mathcal{M}$ now appears in the phase and frequency of the pulsar term (within $\Delta \tau_c$ and $\theta_c$ in equation~\ref{eqn: Res(t) pw freq evo phase E and P}) as well as the amplitude (in equation~\ref{eqn: h+x(t) & h(t)}).  Unlike in the plane-wave monochromatic regime, $\mathcal{M}$ is not in the same combination with $R$ everywhere.  So in principle, a source well described by this model with significant frequency evolution should allow for independent measurements of both of these source parameters.


    \subsection{Fresnel, monochromatic (IIA)}\label{subsec: fresnel, mono}
    
    Here, we provide a new model to the literature that probes a regime we call the `Fresnel monochromatic' regime.  An outline of a derivation for this model is given in Appendix~\ref{app: deriving the timing residual}.  The key to deriving this model is that we keep out to the Fresnel term in the expansion of the retarded time in equation~\ref{eqn:retarded time}.  Conceptually, we account for the additional time delay of a curved gravitational wavefront (as compared to a plane wavefront) along the path of a photon travelling from the pulsar to the Earth, as the example in the right-hand panel of Figure~\ref{fig: 4 regimes / tret contours} shows.  This will slightly alter the phase and frequency of the residual in the pulsar term, which, we will show later, can produce potentially measurable effects.  It is important to realize that this derivation and formalism does not correct for wavefront curvature effects in the geometrical orientation terms like the antenna patterns.  We expect those corrections to introduce curvature effects of order $\mathcal{O}\left(\frac{L}{R}\right)$ in the \textit{amplitude} of the pulsar term, which we do not expect to be measurable.  This is unlike the order $\mathcal{O}\left(\frac{L}{R}\right)$ in the \textit{retarded time}, which can lead to $F\sim \mathcal{O}(1)$ as discussed in Section~\ref{subsec: PW vs Fresnel}.  Due to this higher sensitivity to small corrections in the phase and frequency of the pulsar term, and thanks to how they combine with the Earth term to potentially further produce beats in the timing residual signal, these corrections can influence the timing residual model in a significant way. 
    
    The model we have derived can be expressed as:
    \begin{align}
        \mathrm{Res}(t) &= \hat{p}^i\hat{p}^j E^{\hat{r}\textsc{A}}_{ij} \left(\frac{1}{8}\frac{\lambda_\mathrm{gw}}{c}\frac{R}{c}\frac{1}{\left(1-\left(\hat{r}\cdot\hat{p}\right)^2\right)}\right)^{1/2} \Bigg[  \Big\{C\left(\eta_2\right)-C\left(\eta_1\right)\Big\} h_\textsc{A}\left(\Theta^{'}\right) \ + \ \Big\{S\left(\eta_2\right)-S\left(\eta_1\right)\Big\} h_\textsc{A}\left(\Theta^{'}-\frac{\pi}{4}\right) \Bigg]  \qquad \mathrm{for}\quad \textsc{A} \in [+, \times] , \label{eqn: Res(t) fresnel mono} \\[4pt]
        &\underset{\left(t_0 \ = \ -\frac{R}{c}\right)}{\mathrm{where}}\quad \Theta^{'} \equiv \Theta\left(t+\frac{1}{2}\frac{R}{c}\frac{\left(1-\hat{r}\cdot\hat{p}\right)}{\left(1+\hat{r}\cdot\hat{p}\right)} \right) \equiv \theta_0 + \omega_0\left(t+\frac{1}{2}\frac{R}{c}\frac{\left(1-\hat{r}\cdot\hat{p}\right)}{\left(1+\hat{r}\cdot\hat{p}\right)} \right) , \\
        &\qquad\qquad \ \ \begin{cases}
            \eta_1 \equiv \left(2\frac{R}{\lambda_\mathrm{gw}} \frac{\left(1-\hat{r}\cdot\hat{p}\right)}{\left(1+\hat{r}\cdot\hat{p}\right)}\right)^{1/2} , \\
            \eta_2 \equiv \eta_1 \Big[1+\left(1+\hat{r}\cdot\hat{p}\right)\left(\frac{L}{R}\right)\Big]  ,
        \end{cases}
    \end{align}
    along with equations~\ref{eqn:source/pulsar basis vectors}-\ref{eqn: antenna patterns}, \ref{eqn: gw wavelength}, and~\ref{eqn: h0+x and h0}.  Here, $S$ and $C$ are the Fresnel integrals (see Appendix~\ref{app: deriving the timing residual}).  The Fresnel monochromatic model equation~\ref{eqn: Res(t) fresnel mono} approaches the plane-wave monochromatic model equation~\ref{eqn: Res(t) pw mono} as we would expect in the small Fresnel number limit ($F \rightarrow 0$) and the natural plane-wave limit ($\hat{r}\cdot\hat{p} \rightarrow \pm 1$).  As discussed in Section~\ref{subsec: plane-wave, monochromatic}, both of these effects result in zero residual for these two precise alignments; hence, the natural plane-wave limit will not be of much interest to us going forward in this study.
    
    A very important conclusion here is to notice that in this regime, the degeneracy of the source chirp mass $\mathcal{M}$ and distance $R$ parameters is also broken, like it was in the plane-wave frequency evolution regime.  Here (and in the models in Sections~\ref{subsubsec: fresnel, mono (Asymptotic model)} and~\ref{subsubsec: fresnel, mono (Heuristic model)}), the distance $R$ now appears independently from a combination with the chirp mass $\mathcal{M}$.  So in principle, a source well described by this model with significant Fresnel numbers among the pulsars in the PTA should allow for independent measurements of both of these source parameters.


        \subsubsection{Asymptotic model}\label{subsubsec: fresnel, mono (Asymptotic model)}
        Both $\eta_2, \ \eta_1 \gg 1$ unless they are very near the natural plane-wave limit.  When using this model we will normally not be interested in the natural plane-wave limit, in which case we can replace the Fresnel integrals in equation~\ref{eqn: Res(t) fresnel mono} with their asymptotic expansions (see equation~\ref{equation: Fresnel integrals}).  Doing this and dropping terms of $\mathcal{O}\left(\frac{L}{R}\right)$ in the \textit{amplitude} (not in any phase terms) let's us replace equation~\ref{eqn: Res(t) fresnel mono} with: 
        \begin{equation}
            \mathrm{Res}(t) \approx \frac{F^\textsc{A}}{4\omega_0} \Bigg[  \Bigg\{\sin\bigg(\frac{\pi}{2}\eta_2^2\bigg)-\sin\bigg(\frac{\pi}{2}\eta_1^2\bigg)\Bigg\} h_\textsc{A}\left(\Theta^{'}\right) +  \Bigg\{-\cos\bigg(\frac{\pi}{2}\eta_2^2\bigg)+\cos\bigg(\frac{\pi}{2}\eta_1^2\bigg)\Bigg\} h_\textsc{A}\left(\Theta^{'}-\frac{\pi}{4}\right) \Bigg]   \qquad \mathrm{for}\quad \textsc{A} \in [+, \times], \label{eqn: Res(t) fresnel mono asymptotic}
        \end{equation}
        where now we see the Fresnel formalism solution more closely resembles that of the plane-wave solution equation~\ref{eqn: Res(t) pw mono}, as they both share the same amplitude factor.


        \subsubsection{Heuristic Model}\label{subsubsec: fresnel, mono (Heuristic model)}
        Our study and derivation of the gravitational wave timing residual in both the plane-wave and Fresnel regimes give us an insight into the underlying physics in these models.  In the plane-wave regime, the timing residual depends only on an Earth term and a pulsar term.  We know that a curved wavefront introduces additional time delays in the arrival of the wave at these endpoints (see the right-hand panel of Figure~\ref{fig: 4 regimes / tret contours}).  Therefore, we begin with the plane-wave formalism but modify the retarded time expression at the \textit{pulsar endpoint} to include the higher order correction terms in the retarded time, which come from considering the wavefront curvature.  A priori, we define the following quantity:
        \begin{align}
            \overline{t}_\mathrm{ret} &\equiv t - \frac{L}{c} - \frac{1}{c}\sqrt{R^2 - 2\left(\hat{r}\cdot\hat{p}\right)RL + L^2} , \nonumber \\
            &= t  - \frac{R}{c}   - \left(1-\hat{r}\cdot\hat{p}\right)\frac{L}{c}  - \frac{1}{2}\left(1-\left(\hat{r}\cdot\hat{p}\right)^2\right)\frac{L}{c}\frac{L}{R}  + \ldots .
        \label{eqn: tret model}
        \end{align}
        This expression is nearly identical to the retarded time equation~\ref{eqn:retarded time} evaluated at the source position ($\vec{x}^{'} = R\hat{r}$) and pulsar position ($\vec{x} = L\hat{p}$), except we are explicitly adding in an additional $-\frac{L}{c}$ term that does not appear in that equation.  However, if we define this quantity, then first notice that the first three terms in the above expression recover the same retarded time factor we had for the pulsar term in the plane-wave monochromatic regime equation~\ref{eqn: Res(t) pw mono phase E and P} (which is what we want if this proposed model is to reduce to the plane-wave monochromatic formalism in the appropriate limits).  Secondly, notice that the fourth term then adds back in the Fresnel correction we had previously established in equation~\ref{eqn:retarded time}.
        
        On the basis of this physical interpretation, we therefore write the timing residual in the Fresnel monochromatic regime as:
        \begin{align}
            \overline{\mathrm{Res}}(t) &\equiv \frac{F^\textsc{A}}{4\omega_0} \bigg[ h_\textsc{A}\Big(\Theta_E-\frac{\pi}{4}\Big) - h_\textsc{A}\Big(\overline{\Theta}_P-\frac{\pi}{4}\Big) \bigg]  \qquad \mathrm{for}\quad \textsc{A} \in [+, \times], \label{eqn: Res(t) fresnel mono approx} \\[4pt]
            &\underset{\left(t_0 \ = \ -\frac{R}{c}\right)}{\mathrm{where}}\quad \begin{cases}
                \Theta_E \equiv \Theta\Big(t-\frac{R}{c}\Big) &\equiv \theta_0 + \omega_0 t , \\
                \overline{\Theta}_P \equiv \Theta\big(\overline{t}_\mathrm{ret}\big) &\approx \Theta\Bigg(t - \frac{R}{c}  - \left(1-\hat{r}\cdot\hat{p}\right)\frac{L}{c} - \frac{1}{2}\left(1-\left(\hat{r}\cdot\hat{p}\right)^2\right)\frac{L}{c}\frac{L}{R}\Bigg) , \\
                &\equiv \theta_0 + \omega_0\Bigg(t-\left(1-\hat{r}\cdot\hat{p}\right)\frac{L}{c} - \frac{1}{2}\left(1-\left(\hat{r}\cdot\hat{p}\right)^2\right)\frac{L}{c}\frac{L}{R}\Bigg) ,
                \end{cases}\label{eqn: Res(t) fresnel mono phase E and P}
        \end{align}
        along with equations~\ref{eqn:source/pulsar basis vectors}-\ref{eqn: antenna patterns}, and~\ref{eqn: h0+x and h0}.  The overbar notation used here and below is simply to distinguish that unlike the previous residual models, this has not been analytically derived but rather proposed.
        
        This proposed model has the required properties that it reduces to the plane-wave formalism in the small Fresnel number and natural plane-wave limits.  It takes the general form of the plane-wave formalism (equation~\ref{eqn: Res(t) pw mono}) but with the corrections motivated by the Fresnel regime study.  We tested it numerically and found it to match the results given by the analytic model equation~\ref{eqn: Res(t) fresnel mono} and the asymptotic model equation~\ref{eqn: Res(t) fresnel mono asymptotic} extremely well.  Moreover, we found from a practical standpoint that this formula is easier to study numerically in our Fisher analysis in Section~\ref{sec:fisher matrix analysis}.

        Alternatively, we can decompose equation~\ref{eqn: Res(t) fresnel mono approx} into the sum of a plane-wave part and a correction part due to the curvature of the gravitational wavefront, similar to what was done in~\citet{DF_main_paper}:
        \begin{align}
            \overline{\mathrm{Res}}(t) &\equiv \mathrm{Res}_{(\mathrm{IA})}(t) + \mathrm{Res}_{(\mathrm{IIA,correction})}(t) , \label{eqn: Res(t) fresnel mono approx - alternative expression} \\[4pt]
            &\mathrm{where}\quad \mathrm{Res}_{(\mathrm{IIA,correction})}(t) \equiv \frac{F^\textsc{A}}{4\omega_0} \bigg[ h_\textsc{A}\Big(\Theta_P-\frac{\pi}{4}\Big) - h_\textsc{A}\Big(\overline{\Theta}_P-\frac{\pi}{4}\Big) \bigg] , \nonumber \\
            &\hspace{3.45cm} = A_\mathrm{c(IIA,correction)} \ F^\textsc{A} \ s_{P,\textsc{A}}    \qquad \mathrm{for}\quad \textsc{A} \in [+, \times], \label{eqn: Res(t) fresnel mono - alternative expression correction term} \\[4pt]
            &\qquad\quad \ \begin{cases}
            A_\mathrm{c(IIA,correction)} \equiv \frac{h_\mathrm{c(IIA,correction)}}{4\omega_0} , \\
            h_\mathrm{c(IIA,correction)} \equiv 2 h_0 \sin\left(\frac{\omega_0}{2}\left(1-\left(\hat{r}\cdot\hat{p}\right)^2\right)\frac{L}{c}\frac{L}{R}\right) ,
            \end{cases} \label{eqn: Res(t) fr mono correction characteristic strain and amplitude} \\            
            &\qquad\quad \ \begin{cases}    
            s_{P,+} \equiv \sin\left(2\Theta_P - \frac{\pi}{2} - \frac{\omega_0}{2}\left(1-\left(\hat{r}\cdot\hat{p}\right)^2\right)\frac{L}{c}\frac{L}{R}\right) , \\
            s_{P,\times} \equiv -\cos\left(2\Theta_P - \frac{\pi}{2} - \frac{\omega_0}{2}\left(1-\left(\hat{r}\cdot\hat{p}\right)^2\right)\frac{L}{c}\frac{L}{R}\right) .
            \end{cases}\label{eqn: Res(t) fr mono correction "r" terms}
        \end{align}
        Here, $\mathrm{Res}_{(\mathrm{IA})}(t)$ is the plane-wave monochromatic model from Section~\ref{subsec: plane-wave, monochromatic}, and equation~\ref{eqn: Res(t) fresnel mono - alternative expression correction term} is the correction to the plane-wave regime due to the wavefront curvature from the Fresnel terms.  Note that $\Theta_P$ in equations~\ref{eqn: Res(t) fresnel mono - alternative expression correction term} and~\ref{eqn: Res(t) fr mono correction "r" terms} comes from the IA model equation~\ref{eqn: Res(t) pw mono phase E and P}.  Just like in Section~\ref{subsec: plane-wave, monochromatic}, we can express this correction in terms of its own `characteristic' strain and timing residual amplitudes $ h_\mathrm{c(IIA,correction)}$ and $A_\mathrm{c(IIA,correction)}$.  This provides a very convenient interpretation and comparison of the IA and IIA regimes.  Since the IIA model is identical to the IA model with an additional correction due to the curvature of the gravitational wavefront, and since we now have expressions for the characteristic amplitudes of both the IA timing residual and this correction term, we can define the ratio of the amplitude of the correction to the amplitude of the plane-wave part as:
        \begin{equation}
            \rho \equiv \left|\frac{A_\mathrm{c(IIA,correction)}}{A_\mathrm{c(IA)}}\right| = \left|\frac{h_\mathrm{c(IIA,correction)}}{h_\mathrm{c(IA)}}\right| = \left|\frac{\sin\left(\frac{\omega_0}{2}\left(1-\left(\hat{r}\cdot\hat{p}\right)^2\right)\frac{L}{c}\frac{L}{R}\right)}{\sin\left(\omega_0\left(1-\hat{r}\cdot\hat{p}\right)\frac{L}{c}\right)}\right| .
        \end{equation}
        This ratio compares the relative size of the Fresnel correction in the IIA model to the plane-wave contribution and is a similar quantity again to what~\citeauthor{DF_main_paper} calculated in their work.  If the value of $\rho \sim \mathcal{O}(1)$ or larger, then the Fresnel terms contribute a very significant correction to the otherwise plane-wave monochromatic model.  Together with the Fresnel number itself (equation~\ref{eqn: Fresnel number}), $F$ and $\rho$ are two useful metrics in evaluating the significance of the curvature corrections to our models.


    \subsection{Fresnel, frequency evolution (conjecture) (IIB)}\label{subsec: fresnel, freq evolution (conjecture)}
        
    Given the understanding of the previous timing regimes, we now propose an even more generalized gravitational wave timing residual model, one which includes the Fresnel corrections, allows for frequency evolution, and still recovers all of the previous three regimes in the appropriate limits.  A full mathematical derivation along the lines of what was given in Appendix~\ref{app: deriving the timing residual} would be a useful future project to validate our proposal here.  Our conjecture, however, builds upon the reasoning explained from our insights in Section~\ref{subsubsec: fresnel, mono (Heuristic model)}.  Again, this model corrects only the effects of wavefront curvature in the frequency and phase terms, not in the geometrical orientation terms (as discussed in Section~\ref{subsec: fresnel, mono}).
    
    In the monochromatic formalism, we generalized the plane-wave solution by modifying the retarded time in the pulsar term to include the Fresnel term, which encoded information about the curvature of the wave.  Now we propose doing the same thing with the frequency evolution formalism.  We begin with the plane-wave frequency evolution regime equation~\ref{eqn: Res(t) pw freq evo}, and we modify the pulsar term's retarded time using equation~\ref{eqn: tret model}.  This affects both the orbital frequency of the pulsar term $\overline{\omega}_{0P}$ and the phase $\overline{\theta}_{0P}$.  The model can be expressed compactly as:
    \begin{align}
        \overline{\mathrm{Res}}(t) &= \frac{F^\textsc{A}}{4} \left[ \frac{h_\textsc{A}\Big(\omega_{0E},\Theta_E-\frac{\pi}{4}\Big)}{\omega_{0E}} - \frac{h_\textsc{A}\Big(\overline{\omega}_{0P},\overline{\Theta}_P-\frac{\pi}{4}\Big)}{\overline{\omega}_{0P}} \right]   \qquad \mathrm{for}\quad \textsc{A} \in [+, \times], \label{eqn: Res(t) fr freq evo} \\[4pt]
        &\underset{\left(t_0 \ = \ -\frac{R}{c}\right)}{\mathrm{where}}\quad \begin{cases}
            \omega_{0E} &\equiv \omega_0 , \\
            \overline{\omega}_{0P} &\equiv \omega_0\left[1 + \frac{\left(1-\hat{r}\cdot\hat{p}\right)\frac{L}{c} + \frac{1}{2}\left(1-\left(\hat{r}\cdot\hat{p}\right)^2\right)\frac{L}{c}\frac{L}{R}}{\Delta\tau_c}  \right]^{-3/8} , \\
            \theta_{0E} &= \theta_0 , \\
            \overline{\theta}_{0P} &= \theta_0 + \theta_c\left(1-\left[1+ \frac{\left(1-\hat{r}\cdot\hat{p}\right)\frac{L}{c} + \frac{1}{2}\left(1-\left(\hat{r}\cdot\hat{p}\right)^2\right)\frac{L}{c}\frac{L}{R}}{\Delta\tau_c}\right]^{5/8}\right) , \\
            \Theta_E & = \theta_{0E} + \omega_{0E}t , \\
            \overline{\Theta}_P & = \overline{\theta}_{0P} + \overline{\omega}_{0P}t ,
            \end{cases}\label{eqn: Res(t) fresnel freq evo phase E and P}
    \end{align}
    along with equations~\ref{eqn:source/pulsar basis vectors}-\ref{eqn: antenna patterns}, \ref{eqn: h+x(t) & h(t)}, and~\ref{eqn: coalescence time/angle}.  This model behaves as expected in the previously established limits, recovering all previous regimes.

    And just like we did with the Fresnel monochromatic model in Section~\ref{subsubsec: fresnel, mono (Heuristic model)}, we can also decompose this formula into a plane-wave part and a correction part due to the curvature of the gravitational wavefront:
    
    \begin{align}
        \overline{\mathrm{Res}}(t) &\equiv \mathrm{Res}_{(\mathrm{IB})}(t) + \mathrm{Res}_{(\mathrm{IIB,correction})}(t) , \label{eqn: Res(t) fresnel freq evo - alternative expression} \\[4pt]
        &\mathrm{where}\quad \mathrm{Res}_{(\mathrm{IIB,correction})}(t) \equiv \frac{F^\textsc{A}}{4} \left[ \frac{h_\textsc{A}\Big(\omega_{0P},\Theta_P-\frac{\pi}{4}\Big)}{\omega_{0P}} - \frac{h_\textsc{A}\Big(\overline{\omega}_{0P},\overline{\Theta}_P-\frac{\pi}{4}\Big)}{\overline{\omega}_{0P}} \right]   \qquad \mathrm{for}\quad \textsc{A} \in [+, \times]. \label{eqn: Res(t) fresnel freq evo - alternative expression correction term}
    \end{align}
    Here, $\mathrm{Res}_{(\mathrm{IB})}(t)$ is the plane-wave frequency evolution model from Section~\ref{subsec: plane-wave, freq evolution}, and equation~\ref{eqn: Res(t) fresnel freq evo - alternative expression correction term} is the correction to the plane-wave regime due to the wavefront curvature from the Fresnel terms.  Note that $\omega_{0P}$ and $\Theta_P$ in equation~\ref{eqn: Res(t) fresnel freq evo - alternative expression correction term} come from the IB model equation~\ref{eqn: Res(t) pw freq evo phase E and P}.
    
    Finally, like in the Fresnel monochromatic and plane-wave frequency evolution regimes, there are no degeneracies between any of the source parameters.  In the specific case of the parameters $\mathcal{M}$ and $R$, both appear in combination in the amplitude terms (equation~\ref{eqn: h+x(t) & h(t)}), and both appear independently in the pulsar term phase and frequency functions (frequency evolution effects bring in the chirp mass, while Fresnel corrections bring in the source distance).  So, in principle, a source well described by this model with significant frequency evolution and Fresnel numbers among the pulsars in the PTA should allow for independent measurements of all source parameters.


    \subsection{Regime and model comparisons}
    
\begin{figure}
        \centering
        \includegraphics[width=0.7\linewidth]{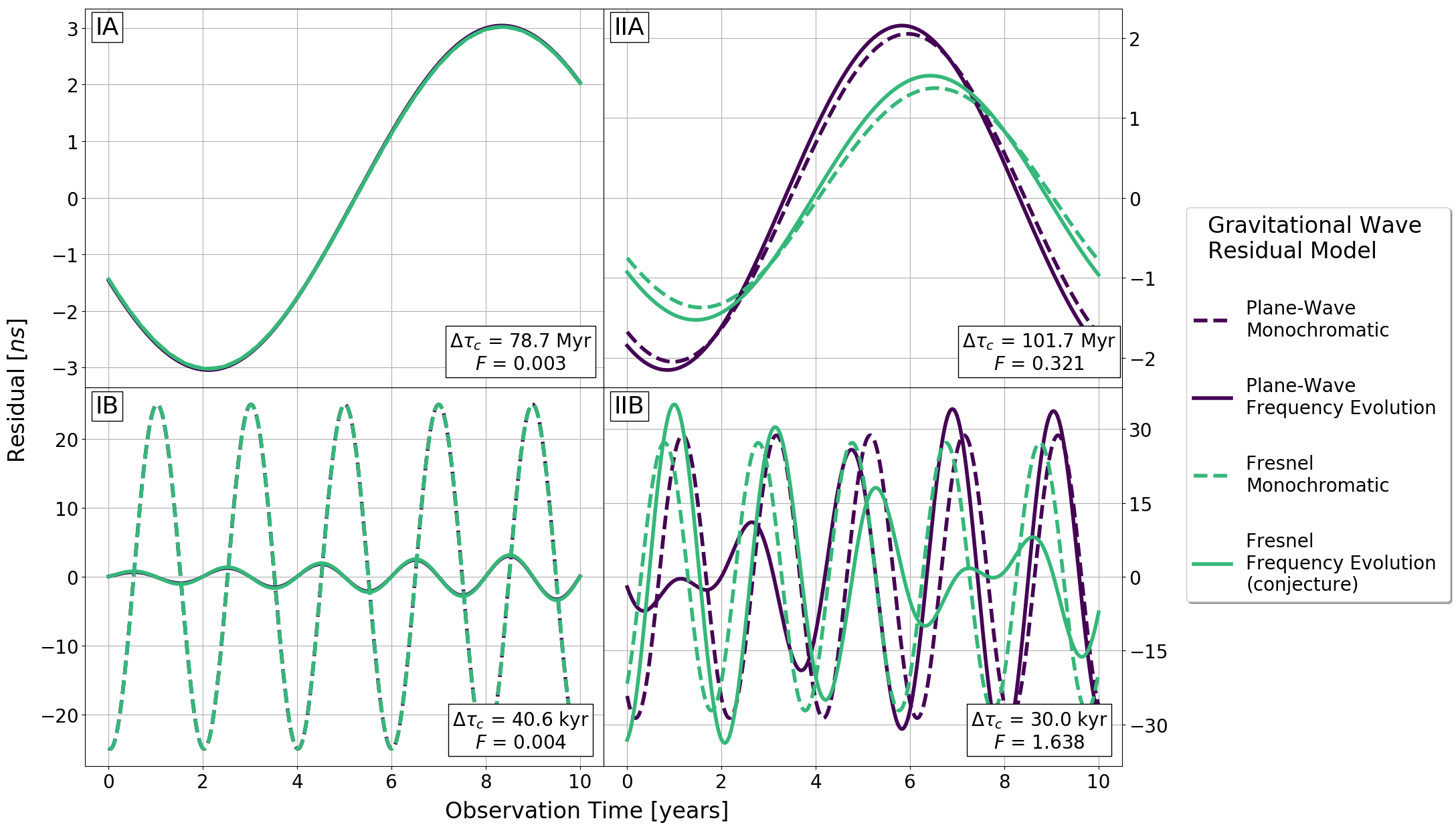}
        \caption{To emphasize regime differences, we plot the gravitational wave timing residuals for a single set of parameters as predicted by each of the four models.  Each panel is labelled by the approximate regime the source is in based on its physical properties (coalescence time and Fresnel number).  In all cases: $\left\{\theta,\phi,\iota,\psi,\theta_0,\theta_p,\phi_p\right\}=\left\{\frac{\pi}{3}, 0, \frac{\pi}{4}, \frac{\pi}{4}, 0, 0, 0\right\} \mathrm{rad}$.  \textbf{(IA)} \underline{Low Fresnel number, high coalescence time}: As a sanity check, we find that all four models do correctly converge on the plane-wave monochromatic regime prediction as expected. Here: $\left\{R,\mathcal{M},\omega_0, L\right\}=\left\{50 \ \mathrm{Mpc}, 1\times10^8 \ M_\odot, 8 \ \mathrm{nHz}, 0.8 \ \mathrm{kpc}\right\}$.  \textbf{(IB)}  \underline{Low Fresnel number, low coalescence time}: The monochromatic models separate from the frequency evolution models since they cannot capture the significant change in the orbital evolution of the source.  Additionally, since the Fresnel number is not significant, the Fresnel models converge on the plane-wave models. Here: $\left\{R,\mathcal{M},\omega_0,L\right\}=\left\{100 \ \mathrm{Mpc}, 5\times10^8 \ M_\odot, 50 \ \mathrm{nHz}, 0.5 \ \mathrm{kpc}\right\}$.  \textbf{(IIA)}  \underline{High Fresnel number, high coalescence time}: The plane-wave models separate from the Fresnel models since they cannot capture the significant effect of the wavefront's curvature.  Additionally, since frequency evolution is not significant, those models converge on the monochromatic model predictions. Here: $\left\{R,\mathcal{M},\omega_0, L\right\}=\left\{50 \ \mathrm{Mpc}, 6\times10^7 \ M_\odot, 10 \ \mathrm{nHz}, 7 \ \mathrm{kpc}\right\}$.  \textbf{(IIB)}~\underline{High Fresnel number, low coalescence time}: Here, the monochromatic models can no longer capture the interesting physics such as the beat frequency due to the frequency evolution of the source, and the plane-wave models cannot reliably account for the additional effects that wavefront curvature introduces.  All four models begin to predict very different results for the same set of parameters due to their respective limitations and underlying assumptions, and the most reliable model becomes the full Fresnel frequency evolution model. Here: $\left\{R,\mathcal{M},\omega_0,L\right\}=\left\{100 \ \mathrm{Mpc}, 6\times10^8 \ M_\odot, 50 \ \mathrm{nHz}, 10 \ \mathrm{kpc}\right\}$.}
        \label{fig: model-regime comparison}
    \end{figure}

    Figure~\ref{fig: model-regime comparison} shows four examples that highlight the differences in these regimes.  It is important to keep in mind the limitations and assumptions under which each of these models are derived when making direct comparisons.  The models can give significantly different and unreliable results if they are being applied to a regime in which they cannot accurately account for that regime's physics.  For example, all four models should give the same predictions for a source with high coalescence time and low Fresnel number since they all can describe the physics of a plane-wave monochromatic source.  However, they should all predict different timing residuals for a source with low coalescence time and high Fresnel number since each of the models (apart from model IIB) have assumptions built into them, which prevent them from describing all of the physics in this scenario (that is IA and IB cannot describe the effects of a curved wavefront, and IIA cannot describe the effects of frequency evolution).
    
    The four model regimes can be further subdivided into different categories given the parameter dependencies of the coalescence time and the Fresnel number (equations~\ref{eqn: coalescence time/angle} and~\ref{eqn: Fresnel number}).  The parameters $\left\{\mathcal{M}, \omega_0\right\}$ control the transition between the monochromatic and frequency evolution regimes, and the parameters $\left\{R, \omega_0, L\right\}$ control the transition between the plane-wave and Fresnel regimes (see the left-hand panel of Figure~\ref{fig: 4 regimes / tret contours}).  Consequently, the orbital frequency parameter has the most direct influence on these regimes - increasing the orbital frequency pushes the models towards \textit{both} the Fresnel and frequency evolution regimes, and vice versa.  Special consideration should be taken of this when studying these models, because systems at higher frequencies will likely be more influenced by both the effects of frequency evolution and the wavefront curvature.  This means that understanding the full Fresnel frequency evolution model is very important, because it has the potential to predict very different timing residuals than the currently used plane-wave frequency evolution model.
    
    We also see that the Fresnel number scales as $L^2$, which means that increasing the pulsar distance is the easiest way to push the residual model into the Fresnel regime.  This gives us motivation to look for and time pulsars at greater distances in our PTAs, since it is the one factor that we have more direct control over in enabling us to explore the Fresnel regime in pulsar timing experiments.  We explore this idea in much greater detail in Section~\ref{sec:results}.


    \subsection{Regime search parameters and degeneracies}\label{subsec:regime degeneracies}
    
    Each of the four model regimes allows for the measurement of different source parameters.  However, certain parameters are degenerate or highly covariant with others in certain regimes, which has motivated us to parametrize the model in specific ways.  Table~\ref{tab: model regime degeneracies} indicates the default parameters we chose in each of our model regimes.

    Starting with the most basic model, the plane-wave monochromatic regime, the source distance $R$, and chirp mass $\mathcal{M}$ are fully degenerate and appear only in the amplitude of the residual (equation~\ref{eqn: Res(t) pw mono}).  We use equations~\ref{eqn: Res(t) pw mono} and~\ref{eqn: h0+x and h0} to introduce the following parameter:
    \begin{equation}
        A_{E,\mathrm{res}} \equiv \frac{h_0}{4 \omega_0} = \frac{\left(G\mathcal{M}\right)^{5/3}}{c^4 R \omega_0^{1/3}} \approx (140 \text{ ns}) \left(\frac{\mathcal{M}}{10^9 M_\odot}\right)^{5/3} \left(\frac{100\text{ Mpc}}{R}\right) \left(\frac{1\text{ nHz}}{\omega_0}\right)^{1/3} ,
    \label{eqn: earth term amplitude}
    \end{equation}
    which is the `Earth term timing residual amplitude', a similar quantity to characteristic amplitudes (equations~\ref{eqn: Res(t) pw mono characteristic strain and amplitude} and~\ref{eqn: Res(t) fr mono correction characteristic strain and amplitude}).  Therefore, in the plane-wave monochromatic regime, this Earth term timing residual amplitude parameter replaces the chirp mass and source distance parameters.  This new parameter also removes the orbital frequency $\omega_0$ from the timing residual amplitude terms in our model.  Therefore, $\omega_0$ is being measured only from its contribution to the time evolution of the phase of the wave (equation~\ref{eqn: Res(t) pw mono phase E and P}), not from its contribution to the timing residual amplitude.
    
    As we discussed in the previous sections, the $R$-$\mathcal{M}$ degeneracy is broken in the other three regimes due to either frequency evolution or Fresnel curvature effects.  However, we found that if we removed the $A_{E,\mathrm{res}}$ parameter from these models and included $R$, $\mathcal{M}$, and $\omega_0$ each separately in the amplitude of the timing residual, it introduced strong covariances between the three parameters, which made it very difficult to measure each of them independently.  This is because the combination of $\frac{\mathcal{M}}{R\omega_0^{1/3}}$ appears in both the Earth and pulsar terms of every timing residual measured, independent of the pulsar being timed, which makes these parameters covariant.  We need the differences caused by frequency evolution and Fresnel corrections that are due to different pulsar distances and locations across the sky to help reduce the covariances between these three parameters.  Therefore, in every regime, we continue to include the Earth term timing residual amplitude $A_{E,\mathrm{res}}$ as its own separate parameter.\footnote{In the two frequency evolution regimes IB and IIB, we wrote the amplitude of the pulsar term as the combination of parameters $A_{P,\mathrm{res}} = A_{E,\mathrm{res}} \left(\frac{\omega_0}{\omega_{0P}}\right)^{1/3}$ for the plane-wave regime, and $A_{P,\mathrm{res}} = A_{E,\mathrm{res}} \left(\frac{\omega_0}{\overline{\omega}_{0P}}\right)^{1/3}$ for the Fresnel regime.}  The other parameters $R$, $\mathcal{M}$, and $\omega_0$ are then measured from the other components of the model where they appear.

    \begin{table}
        \centering
        \caption{Our default source parameters $\vec{s}$ for the timing residual models in each regime, based on the model degeneracies and covariances.  The parameter $A_{E,\mathrm{res}}$ is the `Earth term timing residual amplitude' (equation~\ref{eqn: earth term amplitude}).  In general, frequency evolution allows the direct measurement of chirp mass $\mathcal{M}$ in the frequency $\omega(t)$ and phase $\Theta(t)$ terms in the models, and Fresnel corrections allow for direct measurement of source distance $R$ from the frequency and phase terms.  Fisher matrices with alternative parametrization choices were computed using equation~\ref{eqn: fisher coordinate transformation}.}
        \label{tab: model regime degeneracies}
        \begin{tabular}{ c|c|c } 
        \hline
                 &  Plane-wave  &  Fresnel \\ 
                 \hline
                 Monochrome  &  ${\vec{s}_\mathrm{IA} = \left\{A_{E,\mathrm{res}}, \theta, \phi, \iota, \psi, \theta_0, \omega_0\right\}}$  &  $\vec{s}_{\mathrm{IIA}} = \vec{s}_{\mathrm{IA}} \cup \left\{R\right\}$ \\
                 \hline
                 Frequency evolution  &  $\vec{s}_{\mathrm{IB}} = \vec{s}_{\mathrm{IA}} \cup \left\{\mathcal{M}\right\}$  &  $\vec{s}_{\mathrm{IIB}} = \vec{s}_{\mathrm{IA}} \cup \left\{R,\mathcal{M}\right\}$  \\
                 \hline
        \end{tabular}
    \end{table}

    \section{Fisher Matrix Analysis}\label{sec:fisher matrix analysis}

To study these models, we perform a Fisher matrix analysis to predict how well pulsar timing experiments will be able to measure and constrain the model parameters~(Wittman, unpublished notes~\footnote{Wittman D., (N.D.), Fisher Matrix for Beginners, UC Davis, \url{http://wittman.physics.ucdavis.edu/Fisher-matrix-guide.pdf.}}, \cite{ly_Fisher_tutorial}), similar to~\citet{CC_main_paper}.  This type of analysis is useful in that it allows us to quickly forecast the results of different PTA experiments and look for ways of improving model parameter measurement through the experimental design of the timing experiment itself.  The inverse of the Fisher matrix gives an estimate of the parameter covariance matrix $\mathbfss{F} = \mathbfss{C}^{-1}$.  Therefore, computing a model's Fisher matrix and then inverting it gives us the covariances of all the model parameters, which then tells us how well we should be able to measure these parameters, given the experimental design setup.

Consider our observed timing residual data for every pulsar at every observation time (indexed by $a$) arranged in a $d$-dimensional random variable, $\overrightarrow{\mathrm{Res}} = \left\{ \mathrm{Res}_a : a=1,2,\ldots d \right\}$, and the model parameters in Table~\ref{tab: model regime degeneracies} (indexed by $i,j$) arranged in a $k$-dimensional vector $\vec{X}$.  For this work, we assume that the timing residual data are only the sum of an underlying residual due to a gravitational wave source plus some random noise, that is $\overrightarrow{\mathrm{Res}} = \overrightarrow{\mathrm{Res}}_\mathrm{gw} + \vec{N}$.  The gravitational wave timing residual for every pulsar at every observation is written as a function of $\vec{X}$ by choosing a model from Section~\ref{sec:4 model regimes}, organized into the vector $\overrightarrow{\mathrm{Res}}\left(\vec{X}\right)$.  For our timing data, we assume that timing observations have no covariances and that all timing variances are the same, that is $\bmath{\Sigma} = \mathrm{diag}\left[\sigma^2\right]$.  Here, $\sigma$ is the uncertainty in a timing residual measurement, which in our studies we typically set to $\sigma = 100$ ns for the order of magnitude of present capabilities~\citep{Cordes_2010, Liu_2011, Arzoumanian_2014, NG_11yr_data}.  This assumption is meant to represent a best case scenario, as real pulsar timing experiments model additional correlated red noise when estimating source parameters from their pulsar data sets~\citep{corrnoise_vanHaasteren2013, NG_11yr_data, NG_11yr_cw}.  With this we therefore model the likelihood function as a multivariate Gaussian:
\begin{align}
    \mathcal{L}\left(\overrightarrow{\mathrm{Res}}\mid\vec{X}\right) &= \frac{1}{\sqrt{(2\pi)^d \mathrm{det}(\bmath{\Sigma})}}\exp{\left[-\frac{1}{2}\bigg( \overrightarrow{\mathrm{Res}} - \overrightarrow{\mathrm{Res}}\left(\vec{X}\right) \bigg)^T \bmath{\Sigma}^{-1} \bigg( \overrightarrow{\mathrm{Res}} - \overrightarrow{\mathrm{Res}}\left(\vec{X}\right) \bigg) \right]} , \nonumber \\
    &= \frac{1}{\sqrt{(2\pi)^d}\sigma^d} \prod_a \exp{\left[-\frac{1}{2}\frac{1}{\sigma^2}\left( \mathrm{Res}_a - \mathrm{Res}_a\left(\vec{X}\right) \right)^2 \right]} .
\label{eqn: likelihood function}
\end{align}
Next we use the definition of the Fisher matrix (\citeauthor{ly_Fisher_tutorial}):
\begin{equation}
    \textsf{F}_{ij} \equiv - \Bigg\langle \frac{\partial^2}{\partial X_i \partial X_j} \ln{\mathcal{L}\left(\overrightarrow{\mathrm{Res}}\mid\vec{X}_\mathrm{true}\right)} \Bigg\rangle = \Bigg \langle \left( \frac{\partial}{\partial X_i}\ln{\mathcal{L}\left(\overrightarrow{\mathrm{Res}}\mid\vec{X}_\mathrm{true}\right)}\right) \left(\frac{\partial}{\partial X_j}\ln{\mathcal{L}\left(\overrightarrow{\mathrm{Res}}\mid\vec{X}_\mathrm{true}\right)}\right)  \Bigg\rangle ,
\label{eqn: Fisher definition}
\end{equation}
where the expectation values are taken over the random variable $\overrightarrow{\mathrm{Res}}$, and $\vec{X}_\mathrm{true}$ are the true injected parameters, to find the matrix elements for our likelihood model in equation~\ref{eqn: likelihood function}:
\begin{equation}
    \textsf{F}_{ij} = \sum_a \frac{1}{\sigma^2}\left(\frac{\partial \mathrm{Res}_a\left(\vec{X}_\mathrm{true}\right)}{\partial X_i}\right) \left(\frac{\partial \mathrm{Res}_a\left(\vec{X}_\mathrm{true}\right)}{\partial X_j}\right) .
\label{eqn: Fisher matrix for residuals}
\end{equation}
With this an estimate of the posterior distribution of our parameters $\vec{X}$ is written as a multivariate Gaussian whose covariance matrix is the inverse of our model's Fisher matrix:
\begin{equation}
    p\left(\vec{X}\mid\overrightarrow{\mathrm{Res}}\right) = \sqrt{\frac{\mathrm{det}(\mathbfss{F})}{(2\pi)^k }} \exp{\left[-\frac{1}{2}\left( \vec{X} - \vec{X}_\mathrm{true} \right)^T \mathbfss{F} \left( \vec{X} - \vec{X}_\mathrm{true} \right) \right]} .
\label{eqn: Fisher posterior}
\end{equation}
Therefore knowing the Fisher matrix and computing its inverse will give us an estimate of the parameter covariance matrix, which is our goal.

The main parameters of interest in this model are the source parameters.  However, distances to most pulsars are not known to a high degree of accuracy.  For instance numerous pulsars, including ones timed by NANOGrav, have parallax uncertainties on the order of 100~pc~\citep{NG_11yr_data, pulsar_parallax2019}, which is much larger than the typical gravitational wavelength $\lambda_\mathrm{gw}$ for our sources of interest (see equation~\ref{eqn: gw wavelength}).  The reason why the size of the uncertainty on the pulsar distances is critical for our investigation and results is the topic of Section~\ref{sec:L-wrapping problem}.  Because of these uncertainties previous studies have thought to include the pulsar distances as free parameters in addition to the source parameters in their analyses~\citep{CC_main_paper, GWastro_Lee2011}.  This allows us to use the gravitational wave data to also help measure the distances to pulsars in our PTA.

When including pulsar distances as model parameters, the Fisher matrix takes a symmetric block-matrix form, separating into a source parameter only symmetric matrix $\mathbfss{F}^\mathbfss{S}$, a pulsar distance parameter only symmetric matrix $\mathbfss{F}^\mathbfss{L}$, and a matrix with cross terms $\mathbfss{F}^{\mathbfss{SL}}$. Dividing the model parameters into `source' parameters and `pulsar distance' parameters, $\vec{X} = \left[\vec{s}, \vec{L}\right]$, equation~\ref{eqn: Fisher matrix for residuals} then becomes:
\begin{align}
    \begin{cases}
        \textsf{F}^\textsf{S}_{ij} &= \sum\limits_a \frac{1}{\sigma^2}\left(\frac{\partial \mathrm{Res}_a\left(\vec{X}_\mathrm{true}\right)}{\partial s_i}\right) \left(\frac{\partial \mathrm{Res}_a\left(\vec{X}_\mathrm{true}\right)}{\partial s_j}\right)  \\[8pt]
        \textsf{F}^\textsf{SL}_{ij} &= \sum\limits_a \frac{1}{\sigma^2}\left(\frac{\partial \mathrm{Res}_a\left(\vec{X}_\mathrm{true}\right)}{\partial s_i}\right) \left(\frac{\partial \mathrm{Res}_a\left(\vec{X}_\mathrm{true}\right)}{\partial L_j}\right)  \\[8pt]
        \textsf{F}^\textsf{L}_{ij} &= \sum\limits_a \frac{1}{\sigma^2}\left(\frac{\partial \mathrm{Res}_a\left(\vec{X}_\mathrm{true}\right)}{\partial L_i}\right)^2 \delta_{ij} 
    \end{cases} \qquad \longleftrightarrow \qquad \mathbfss{F} = \left[\begin{array}{c|ccc}
            \\[-5pt]
            \mathbfss{F}^\mathbfss{S}                         & \cdots & \mathbfss{F}^{\mathbfss{SL}}  & \cdots \\[8pt] \hline
            \vdots                                & \ddots &                        & 0 \\
            \left(\mathbfss{F}^{\mathbfss{SL}}\right)^T  &        & \mathbfss{F}^\mathbfss{L}          & \\
            \vdots                                & 0      &                        & \ddots
        \end{array}\right] .
\label{eqn: Fisher block matrix}
\end{align}

In our work, we computed the Fisher matrix and resulting covariance matrix in terms of the parameters $\vec{s}$ indicated within each regime in Table~\ref{tab: model regime degeneracies}, and the parameters $\vec{L} = \left[ L_1, L_2, \ldots \right]$ for each pulsar in our array.  However, sometimes it was useful to compute the Fisher matrix for a different parametrization choice, $\vec{X}^{'}$ (for example, parametrizing the model in the \textit{log} of a parameter).  We computed the new transformed Fisher matrix through the Jacobian $\mathbfss{J}$ of our transformation:
\begin{align}
    &\textsf{F}^{'}_{ij} = \textsf{J}_{mi} \textsf{J}_{nj} \textsf{F}_{mn} \quad \longleftrightarrow \quad \mathbfss{F}^{\bmath{'}} = \mathbfss{J}^T \mathbfss{F} \mathbfss{J} , \label{eqn: fisher coordinate transformation} \\
    &\mathrm{where}\quad \textsf{J}_{ij} \equiv \frac{\partial X_i}{\partial X^{'}_j} . \nonumber
\end{align}

For our Fisher analysis, we used the PYTHON SymPy package to first symbolically write and manipulate our models from Section~\ref{sec:4 model regimes}.  Inverting the Fisher matrix was accomplished through singular value decomposition (SVD).  We checked that the condition number of each Fisher matrix was below $10^{14}$ before performing the inversion~\citep{condition_number_source}.  This requirement helped to ensure that the SVD procedure accurately calculated the inverse matrix - if the condition number exceeded this value, then we did not calculate the covariance matrix.  As a final check for accuracy the computed covariance matrix was multiplied against its original Fisher matrix, and the result subtracted from the identity matrix.  This was checked against an `error threshold' matrix defined by:
\begin{equation}
    \mathbfss{I} - \mathbfss{F}\mathbfss{C} < \epsilon \left[\begin{array}{cc} 1 & \cdots \\ \vdots & \ddots \\ \end{array}\right] ,
\end{equation}
where $\epsilon$ was our `error threshold.'  For all results presented in this paper, $\epsilon \leq 0.01$ (and in almost all cases $\leq 10^{-4}$).

    \section{Pulsar Distance Wrapping Problem}\label{sec:L-wrapping problem}

A crucial problem in parameter estimation from continuous gravitational wave pulsar timing residuals is the `pulsar distance wrapping problem' \citep{CC_main_paper, Ellis_2013}.  The pulsar distance $L$ affects the phase of the timing residual terms (see equations~\ref{eqn: Res(t) pw mono phase E and P}, \ref{eqn: Res(t) pw freq evo phase E and P}, \ref{eqn: Res(t) fresnel mono phase E and P}, and~\ref{eqn: Res(t) fresnel freq evo phase E and P}).  Since the phase wraps around the interval $[0,2\pi)$, if the pulsar distance is an unrestricted free parameter, it can become difficult or even impossible to recover in parameter estimation.\footnote{Interestingly, it is also because the phase wraps around $2\pi$ that we can even try to measure $R$ from the Fresnel formalism in the first place!  Corrections of $\mathcal{O}\left(\frac{L}{R}\right)$ to the amplitude of the timing residual model would likely be impossible to measure.  However, corrections of $\mathcal{O}\left(\frac{L}{R}\right)$ that appear in the \textit{phase} they can become appreciable over the $2\pi$-interval through the Fresnel number.  If the phase did not cycle on the interval $[0,2\pi)$, then the Fresnel term in the retarded time equation~\ref{eqn:retarded time} would remain a much smaller correction to the plane-wave approximation and not contribute in a significant way to the overall timing residual.  This idea was briefly discussed in Sections~\ref{subsec: PW vs Fresnel} and~\ref{subsec: fresnel, mono}.}  Consider, for example, the two monochromatic regimes IA and IIA.  Given their pulsar term phase dependence $\Theta_p$ on $L$ (equations~\ref{eqn: Res(t) pw mono phase E and P} and~\ref{eqn: Res(t) fresnel mono phase E and P}), any $L^{'} = L \pm \Delta L_n$ where:
\begin{equation}
\Delta L_n \equiv \begin{cases}
    n \frac{\lambda_\mathrm{gw}}{\left(1-\hat{r}\cdot\hat{p}\right)} ,  &\quad\text{(Plane-Wave, monochromatic)} \\[10pt]
    -\left(\frac{R}{\left(1+\hat{r}\cdot\hat{p}\right)} + L\right) + \sqrt{ \left(\frac{R}{\left(1+\hat{r}\cdot\hat{p}\right)} + L\right)^2 + n\frac{2\lambda_\mathrm{gw}R}{\left(1-\left(\hat{r}\cdot\hat{p}\right)^2\right)}} , &\quad\text{(\textit{Heuristic} Fresnel, monochromatic)}
\end{cases}
\label{eqn:Delta L in monochromatic regimes}
\end{equation}
for $n \in \mathbb{Z}$ will give the same timing residual.  We will refer to $\Delta L_1$ for the \textit{plane-wave monochromatic case} as `one wrapping cycle' in this paper.  Note, assuming $\hat{r}\cdot\hat{p} \neq \pm 1$, the Fresnel $\Delta L_n$ reduces to the plane-wave $\Delta L_n$ in the limit $R\rightarrow\infty$ as expected.  The frequency evolution regimes do not have an easily defined quantity like this because their frequencies are time-dependent.

These relations give us a helpful proxy of how small our observational uncertainty on the pulsar distances $\sigma_L$ need to be if we are to include pulsar distances as free parameters in our models.  Namely $\sigma_L \sim \Delta L_1$, one wrapping cycle, which means that the uncertainty on our pulsar distance measurements needs to be on the order of the wavelength of the gravitational wave $\lambda_\mathrm{gw}$ that we are trying to measure.  For typical sources of interest ($\omega_0 \sim$ 1 nHz to 100 nHz), the $\lambda_\mathrm{gw}$ can range between tens of parsecs to subparsec distances, which imposes a very strong experimental challenge by today's standards.

Note additionally that the geometric scaling factors in equation~\ref{eqn:Delta L in monochromatic regimes} can either help or hurt us, depending on if the pulsar happens to be more aligned or anti-aligned with the source.  Considering the plane-wave monochromatic pulsar wrapping distance $\Delta L_n$, if our pulsar happens to be more aligned with our source (within 90\degr, \ $0 < \hat{r}\cdot\hat{p} < 1$), then the condition relaxes and we do not need as tight of an experimental uncertainty on the pulsar's distance measurement a priori.  But if our pulsar happens to be more anti-aligned with our source (more than 90\degr, \ $-1 < \hat{r}\cdot\hat{p} < 0$), then we require even more precise distance measurements (the smallest $\Delta L_n$ being half a gravitational wavelength).  In Table~\ref{tab: pta reference} we provide for reference the values of the wrapping cycle $\Delta L_1$ proxy (from the plane-wave monochromatic formula) for each pulsar in our PTA with source 1 from Table~\ref{tab: source parameters reference}.

In practice, we found that this problem did not pose a significant issue in our Fisher matrix analyses when using the plane-wave regime models IA and IB but did pose a significant issue when using with the Fresnel regime models IIA and IIB.  In the case of the Fresnel regime models, if no restrictions were placed on the pulsar distance measurements at all, then parameter estimation was not possible because the Fisher matrices could not be accurately inverted (their condition number exceeding $10^{14}$).  Therefore, we addressed this problem in our simulations by adding an uncorrelated Gaussian prior to our knowledge of the pulsar distances in our Fisher matrix~(Wittman, unpublished notes).  Adding $\mathbfss{F}^{\mathbfss{L}\bmath{'}} = \mathbfss{F}^\mathbfss{L} + \left(\mathbfss{C}^{\mathbfss{L}}\right)^{-1}$ to the pulsar distance sub-matrix in equation~\ref{eqn: Fisher block matrix}, where:
\begin{equation}
    \mathbfss{C}^{\mathbfss{L}} = \text{diag}\left(\sigma_L^2\right) ,
\label{eqn:pulsar uncertainty}
\end{equation}
represents pulsar distance measurement constraints placed on our experiments, each pulsar with its own distance uncertainty $\sigma_L$.  In practice, these uncertainties would likely come from electromagnetic observations of the pulsars.

In the case of the plane-wave regime models, we did find it possible to accurately invert the Fisher matrix and recover the source parameters without the need for this additional pulsar distance prior.  Therefore, we conclude that the cause for this is the introduction of the distance parameter $R$ in the Fresnel regimes, which is not in the plane-wave regimes.  As we discuss in detail in Section~\ref{sec:results} the parameter $R$ is the most difficult parameter to measure from our models.  This parameter enters the Fresnel models IIA and IIB in the combination $L/R$; therefore, it is highly sensitive to the pulsar distance wrapping problem and requires the additional pulsar distance prior in order to constrain the measurement.

    \section{Results}\label{sec:results}

For this study, we used the PTA in Table~\ref{tab: pta reference}, an array of 40 of the pulsars from NANOGrav's PTA in~\citet{NG_11yr_data}.  In our simulations we timed each of these pulsars for an observation time of 10 yr, with a timing cadence of 30 observations/year (unless specified otherwise).  Each of these pulsars was timed at the exact same times, and all of the timings were evenly spaced over the observation period.  Additionally, we simulated timing uncertainty $\sigma$ uniformly in all of the timing residuals (see equation~\ref{eqn: Fisher matrix for residuals}) and distance uncertainties $\sigma_L$ for each of the pulsars (see equation~\ref{eqn:pulsar uncertainty}).  Finally, in order to allow us to control the baseline distances to all of the pulsars in our PTA, we introduced a PTA distance `scale factor' term.  Therefore, a scale factor of 1 indicates that all of the pulsars in the PTA have their standard distances given by the $L$ column in Table~\ref{tab: pta reference}, and at most, we increased the scale factor to 7 (because that would place our farthest pulsar \#~40 at roughly the distance of the Large Magellanic Cloud).

In a real pulsar timing experiment, there are other physical effects that create timing residuals, which must also be included in a complete timing residual model~\citep{handbook_pulsar_astro, GWastro_Lee2011, NG_11yr_data}.  For simplicity we wanted to focus solely on the effects of the gravitational waves, without simultaneously modelling other timing residual sources, to help ensure that the gravitational wave parameters would not be degenerate with other model parameters.  The sources we consider in this work have high enough frequencies (with $\omega_0 > 10$ nHz) that we found that their overall signal-to-noise ratio was not significantly impacted by subtracting a quadratic fit to the timing residual (which would roughly approximate of the impact of fitting a more complete timing model).  A simple analysis of this is provided in the supplementary material.  This is in line with \citet{toa_hazboun2019}, who provide a much more in-depth analysis of the general sensitivity of pulsar timing to gravitational waves.

In our work, we use the coefficient of variation (CV) of a given model parameter $x$ as a proxy for that parameter's measureability:
\begin{equation}
    \mathrm{CV}_x \equiv \frac{\text{Standard Deviation of } x}{\text{Expectation Value of } x} = \begin{cases} \frac{\sigma}{\mu} ,  &\quad\text{Normal Distribution in $x$}  \\[8pt]
    \sqrt{e^{\sigma^2 \ln(10)^2} - 1} ,  &\quad\text{Log}_{10}\text{-Normal Distribution in $x$}
    \end{cases}
\label{eqn: CV}
\end{equation}
where $\mu$ and $\sigma$ are the distributions' parameters.  For a normally distributed $x$ parameter, $\mu$ and $\sigma$ are also the distribution's mean and standard deviation.  For a lognormally distributed $x$ parameter, the CV depends only on the lognormal $\sigma$ parameter.  Equation~\ref{eqn: CV} is effectively the fractional error of a given parameter $x$, or a measure of the dispersion of that parameter's distribution.  Once we have the covariance matrix of our parameters from our Fisher analysis, we take the $\sigma$ values from that matrix and compute the CVs based on how each respective parameter is distributed (in the case of the regular normal distribution, $\mu$ is the parameter value we inject into the simulation).  All parameters within a model are normally distributed in a Fisher matrix approximation (see again equation~\ref{eqn: Fisher posterior}), so we need only the second line of equation~\ref{eqn: CV} if we parametrize a model parameter $x$ as $\log_{10}(x)$.  Therefore a parameter with a small CV suggests to us that it would be measurable from the experimental setup, while a CV of order unity or larger would suggest an unmeasurable parameter.


    \subsection{Measuring source distance}\label{subsec: Measuring Source Distance}
    
    As a point of reference, we used Source 1 in Table~\ref{tab: source parameters reference} as a fiducial reference.  This is a  source worth considering because it has a low $\Delta \tau_c$ and is therefore strongly chirping, has a high $A_{E,\mathrm{res}}$ even when the source is at $R=100$ Mpc (which is good if our timing uncertainty $\sigma \sim 100$ ns), and has a $\lambda_\mathrm{gw} \sim 1$ pc, which sets the length scale of the typical wrapping cycle.


        \subsubsection{Figure-of-merit (F.O.M)}\label{subsubsec: Figure-of-Merit}
        One of the primary goals of this work was to determine how well we could measure the source distance $R$ from the Fresnel corrections in our pulsar timing models.  As a starting question we asked, `What is the most important factor in governing the recoverability of the source distance parameter?'  The main takeaway we found is that fundamentally, in order to exploit the Fresnel corrections to our model to measure $R$, our precision is governed by the tightest we can constrain the pulsar distance uncertainties $\sigma_L$ on the \textit{farthest} pulsars in our PTA.
        
        To arrive at this conclusion, we first simulated two PTA experiments, each with a selection of 39 `near' pulsars ($L < 1$ kpc) and a single `distant' pulsar (using the F.O.M. $L$ values indicated in Table~\ref{tab: pta reference}).  Source 1 was placed at a very nearby distance of $R=10$ Mpc.  In the first, case we simulated a prior distance constraint only on the single distant pulsar to within its own wrapping cycle $\sigma_L = \Delta L_1$, while all near pulsars were unconstrained.  In the second case, we did the opposite, placing no prior constraint on the position of the single distant pulsar and giving all nearby pulsars priors equal to their wrapping cycles.  For both scenarios, we built up the size of the PTA by first simulating the timing array with only the single distant pulsar \#~40.  Then for every subsequent simulation, we added in one additional nearby pulsar, starting at pulsar \#~1.  For every simulation, we recorded the PTA's ability to measure the source distance parameter $R$ by calculating the $\mathrm{CV}_R$ value.
        \begin{figure}
            \centering
            \includegraphics[width=0.6\linewidth]{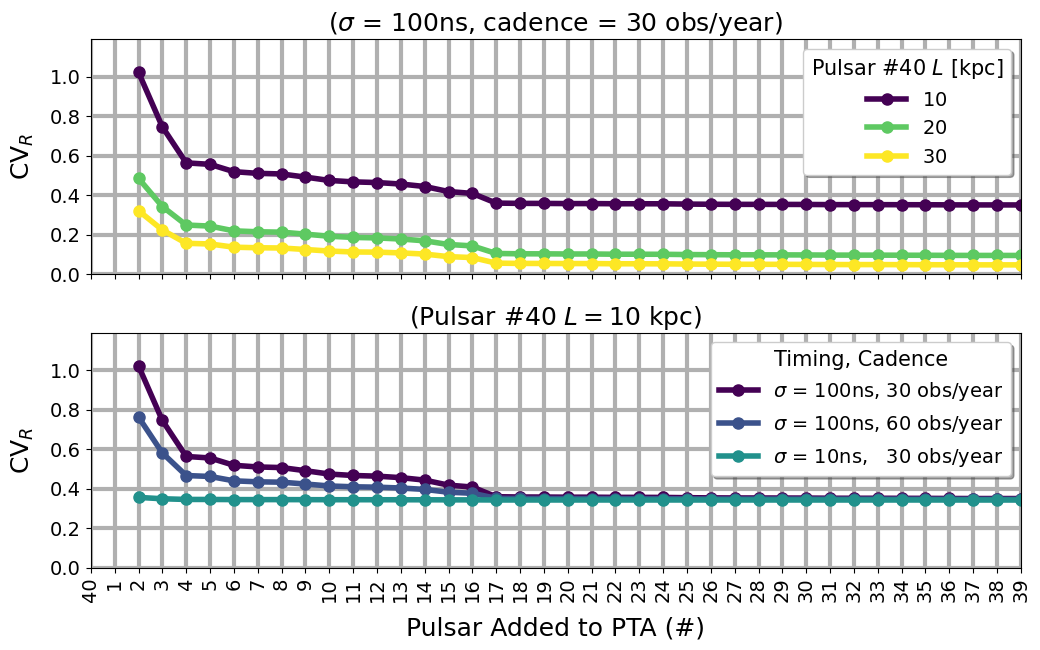}
        \caption{Two figure-of-merit studies investigating what is needed in a PTA to measure the source distance $R$ from Fresnel corrections.  See Table~\ref{tab: pta reference} for the pulsars and their F.O.M $L$ values that were used in these calculations.  Source 1 from Table~\ref{tab: source parameters reference} is simulated at a distance $R=10$ Mpc (therefore, $A_{E,\mathrm{res}}=21 \ \mu$s).  Note that because this is simply meant to be a figure-of-merit study to understand the general behavior of changing the pulsar and PTA parameters, we purposefully simulated a very near and loud source.  In both panels, we first simulated the PTA with only the single `distant' pulsar (pulsar \#~40), placed at the distances indicated, with a distance uncertainty constraint of a wrapping cycle $\sigma_L = \Delta L_1$.  For every subsequent simulation, we added in one additional `nearby' pulsar ($L < 1$ kpc) with no distance prior constraints, starting with pulsar \#~1 through pulsar \#~39.  In each simulation we calculated the $\mathrm{CV}_R$ value for that experiment.  In all cases, we found that we needed a minimum PTA size of 3 in order to accurately invert our Fisher matrix using SVD.  Interestingly, beyond 18 pulsars in the PTA, adding additional pulsars did not seem to improve the recoverability of $R$.  These results were computed using the IIB model (Section~\ref{subsec: fresnel, freq evolution (conjecture)}).}
            \label{fig: FOM - 39 near no-prior, 1 far prior}
        \end{figure}

        \begin{figure}
            \centering
            \includegraphics[width=0.6\linewidth]{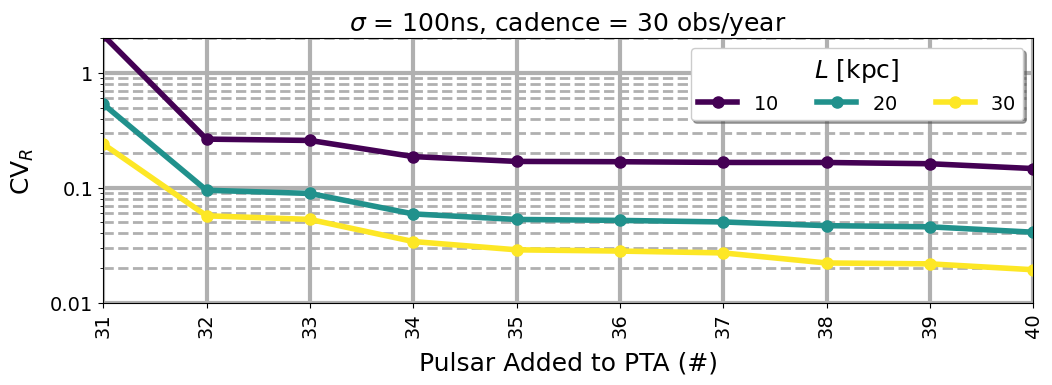}
        \caption{A figure-of-merit study that shows the improvement of measuring $R$ as additional constrained distant pulsars are added to a PTA.  Here, the base PTA consists of nearby pulsars \#~1-30 from Table~\ref{tab: pta reference} using their F.O.M $L$ values, with no prior knowledge distance constraints.  Pulsars \#~31-40 are then added consecutively into the PTA, each with the $L$ distance indicated in the figure, and each with an uncertainty prior of its own wrapping cycle $\sigma_L = \Delta L_1$.  We used the same source and source distance $R$ as in Figure~\ref{fig: FOM - 39 near no-prior, 1 far prior}.  These results were computed using the IIB model (Section~\ref{subsec: fresnel, freq evolution (conjecture)}). }
            \label{fig: FOM - 30 near no-prior, 10 far prior}
        \end{figure} 
        
        Only the first scenario succeeded in producing a reliable measurement of the source distance - the PTA with one well-constrained distant pulsar and many unconstrained nearby pulsars.  Results for this scenario are shown in the top panel of Figure~\ref{fig: FOM - 39 near no-prior, 1 far prior}.  This experimental setup shows a marked improvement in the $\mathrm{CV}_R$ value as the distance to the furthest pulsar is increased (while still holding its $\sigma_L$ fixed).  However, in the opposite scenario when all nearby sources were known to within their wrapping cycles but the single distant pulsar was unconstrained, all measurements of $\mathrm{CV}_R$ were well above unity.  Therefore, even when increasing the distance of the distant pulsar in the array, we found that prior knowledge only on the nearby pulsars was not enough to allow us to measure the value of the source distance from the Fresnel corrections.
        
        In summary, even if the distances of many nearby pulsars are known to a great deal of precision, we cannot recover $R$ unless we can strongly constrain the distance to the furthest pulsars in the array.  In fact, that is \textit{all} we need in principle - we do not even need to have distance measurement constraints on the nearest pulsars in the array a priori.  Our conclusion for the reason that these two F.O.M. studies gave us these results is because nearby pulsars will likely have negligible Fresnel numbers, meaning that we cannot probe their Fresnel corrections to measure $R$.  Even if these nearby pulsars have very well-constrained distance measurements, it is not enough since the actual magnitude of the Fresnel corrections is too small for these pulsars.  Therefore, in order to actually exploit knowledge coming from the Fresnel corrections in our models, we need to have distant pulsars in the array, which have sufficiently high Fresnel numbers.  But due to the pulsar distance wrapping problem, if these distant pulsars are not well-constrained, then we once again lose the ability to measure the source distance $R$.
        
        We also explored the effects of increased observation cadence and improved timing uncertainty $\sigma$ on measuring $R$.  Using again a single constrained distant pulsar and 39 nearby pulsars, the bottom panel of Figure~\ref{fig: FOM - 39 near no-prior, 1 far prior} shows that if the PTA is small in size, a higher timing cadence does help, but timing resolution makes the largest difference.  However, what is interesting is that regardless of the experimental timing uncertainty or cadence, all three schemes shown produce the same level of accuracy at recovering $R$ once our PTA is beyond about 17 pulsars in size.  This means that a small PTA with very well-timed pulsars is good, but if high timing precision cannot be achieved, then adding more pulsars to the PTA will help just as much.  Similarly, cutting a PTA in half and doubling its observation cadence will produce roughly the same results.  
        
        More generally, we observed during our studies that the $\mathrm{CV}$ of the model parameters tended to follow the power law $\mathrm{CV}\propto \left[\sigma / \sqrt{\mathrm{cadence}}\right]^p$.  The parameters $A$, $\iota$, $\psi$, and $\theta_0$ very strongly followed this law with $p=1$, and $\omega_0$ followed with $p\simeq 1$ in most cases.  For the remaining model parameters the $\mathrm{CV}$ still appeared to behave as a univariate function of $\sigma / \sqrt{\mathrm{cadence}}$, but the power-law dependence was not as strong as with the previous parameters.  In most observed cases, the power law could be fit with differing values of $p<1$ (based on the PTA setup and source) for the parameters $\theta$, $\phi$, and $\mathcal{M}$, and still held valid over a large part of the $\sigma - \mathrm{cadence}$ parameter space.  However, we found that the $\mathrm{CV}_R$ was only a very weak function of this variable.
        
        As a final study, since we found that nearby pulsars do not help with the measurement of $R$, we therefore investigated the effect of adding additional constrained \textit{distant} pulsars into the PTA.  In Figure~\ref{fig: FOM - 30 near no-prior, 10 far prior}, we started the initial PTA with pulsars \#~1-30 from Table~\ref{tab: pta reference} (again using their F.O.M. $L$ distances) and then consecutively added the final 10 (each of these with the distances indicated in Figure~\ref{fig: FOM - 30 near no-prior, 10 far prior}).  Again for this study, nearby pulsars were unconstrained, while each added distant pulsar was simulated with its own wrapping cycle prior $\sigma_L = \Delta L_1$.  As suspected, we can see that adding additional well-constrained distant pulsars improves the measurement of $R$.  In a separate study, we also found that if only the first of the distant pulsars was given a wrapping cycle prior constraint, then adding additional \textit{unconstrained} distant pulsars did not improve the measurement of $R$ notably.  So, it appears that the only way to improve and actually probe the source distance from Fresnel corrections really is to have a PTA with numerous distance pulsars, whose distances are very well measured.


        \subsubsection{Distant sources}\label{subsubsec: distant sources}
        Based on our findings, a direct measurement of $R$ given current and near future standards would be difficult, unless perhaps the source happened to be very nearby, such as in the Virgo Cluster.  But for consideration of the direction of future technologies, distant sources on the order of $100$ Mpc to $1$ Gpc would require significant improvements in our PTA.  From our figure-of-merit studies, we know that nearby pulsars do not contribute their Fresnel corrections towards improving the measurement of $R$ - for this we need many distant pulsars in the array with $\sigma_L$ uncertainties on the order of their wrapping distances.  Therefore, we focused our investigation on what scenarios would give us good measurements of $R$ and how this might be useful in the future.
        
\begin{figure}
            \centering
              \begin{subfigure}[t]{0.42\linewidth}
              \centering
                \includegraphics[width=1\linewidth]{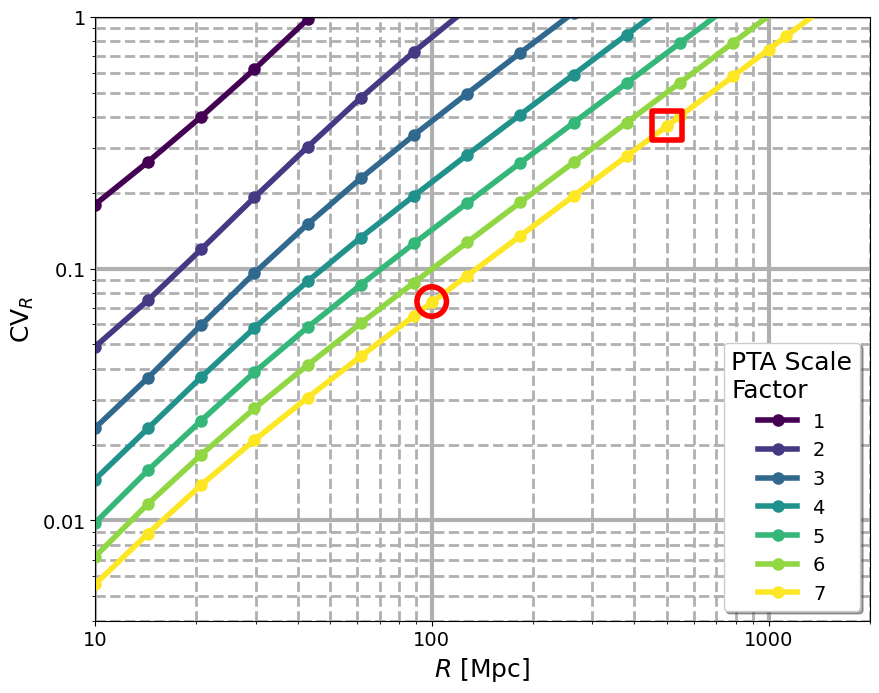}
              \end{subfigure}
              \hfill
              \begin{subfigure}[t]{0.42\linewidth}
              \centering
              \includegraphics[width=1\linewidth]{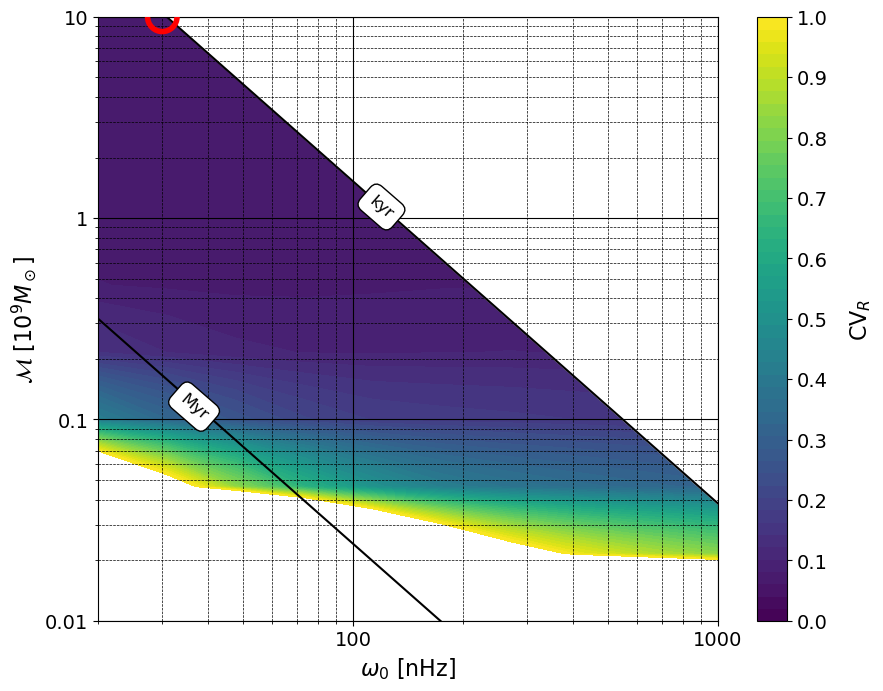}
              \end{subfigure}
        \caption{Source distance measurement as a function of distance and as a function of source intrinsic parameters $\mathcal{M}$ and $\omega_0$.  In both panels, the timing uncertainty $\sigma=100$ ns, the pulsar distance uncertainty $\sigma_L = 1$ pc.  The red circle indicates corresponding points between the two graphs, and the red square corresponds to the plot found in the supplementary material.  Note, a scale factor of $7$ puts the furthest pulsar in our PTA at roughly the distance of the Large Magellanic Cloud.  These results were computed using the IIB model (Section~\ref{subsec: fresnel, freq evolution (conjecture)}).  \textbf{(Left)} Here, we measure $R$ from Source 1 in Table~\ref{tab: source parameters reference}.  Note that $\sigma_L = 1 \ \mathrm{pc}\ \sim \lambda_\mathrm{gw}$ for this source.  \textbf{(Right)} These sources have the same angular parameters as our fiducial Source 1 (indicated in Table~\ref{tab: source parameters reference}) and are set at $R=100$ Mpc. The PTA has a scale factor of $7$.  Contours of coalescence time $\Delta \tau_c$ (equation~\ref{eqn: coalescence time/angle}) are indicated, with a cut along $\Delta \tau_c = 1$ kyr due to assumption~\ref{as: t_obs << tau_c}.  Bias leans towards lower frequencies, where using a fixed $\sigma_L$ value will benefit sources with larger wrapping cycles as compared to smaller ones (see Section~\ref{subsubsec: distant sources}).}
        \label{fig: s1 scaled PTA / M-omega0 space sf7 PTA}
        \end{figure}

        The left-hand panel of Figure~\ref{fig: s1 scaled PTA / M-omega0 space sf7 PTA} shows how increasing the distances of all of the pulsars with the distance scale factor improves the measurement of $R$.  In effect, we will need pulsars across the entire span of our Galaxy, and ideally out to the Large and Small Magellanic Clouds, with distance uncertainties on the order of $\Delta L_1$ or $\lambda_\mathrm{gw}$ in order to probe source distances beyond $100$ Mpc.  Since the Fresnel number scales as $L^2$, increasing the scale factor of our PTA yields a marked effect towards recovering $R$.  If, however, the pulsar distance uncertainties also scaled as $L^2$, which we would expect for measurements of pulsar distances made using parallax, then we found that these effects effectively cancelled each other out, resulting in no improvement in the recovery of $R$ even with larger PTAs.  Therefore, it is crucial that as the pulsar distance in our PTAs increases, their uncertainties continue to remain of the order of the wrapping cycle (a quantity that is \textit{independent} of the pulsar's distance).
        
        Sources with certain physical properties will also be favoured when measuring Fresnel corrections, as seen in the right-hand panel of Figure~\ref{fig: s1 scaled PTA / M-omega0 space sf7 PTA}.  In terms of intrinsic parameters, strongly chirping sources with coalescence times $\Delta \tau_c \sim \mathrm{kyr} - \mathrm{Myr}$ will yield the best measurements of $R$ through Fresnel corrections.  The bias we see in this figure leans towards lower frequency sources, largely because these sources will have larger wrapping cycles and therefore require less pulsar distance precision than higher frequency sources (see equations~\ref{eqn:Delta L in monochromatic regimes} and~\ref{eqn: gw wavelength}).  Therefore, keeping $\sigma_L = 1$ pc pinned means that the lower frequency sources benefit from a higher degree of precision than the higher frequency sources.
        
        While our focus so far has been on the recovery of the distance parameter, we also looked at the measurement of all source parameters.  As an example, Source 1 was considered at $R = 500$ Mpc.  Again, this is a loud source and $\mathrm{CV}_{A_{E,\mathrm{res}}} = 0.0097$ for our simulated $\sigma = 100$ ns timing uncertainty.  In this case, $\mathrm{CV}_R = 0.37$, while all remaining parameters had their respective values of $\mathrm{CV} \leq 0.018$.  Sky angles tend to be measured well for the reasons discussed in Section~\ref{subsec: Source Localization}.  Plus as this is a highly chirping source, chirp mass and frequency are well measured.  A full plot of the distributions and covariances for the parameters in this example can be found in the supplementary material.
        
        For this particular example, we found that improving the timing uncertainty by an order of magnitude to $\sigma = 10$ ns improved the indicated $\mathrm{CV}$ values by one order of magnitude for the parameters $\left\{A_{E,\mathrm{res}}, \iota, \psi, \theta_0  \right\}$, while all remaining parameter $\mathrm{CV}$ values remained the same.  In general, we found that these four parameters were most sensitive to timing accuracy, while mostly insensitive to pulsar distance accuracy.  While the measurement of $R$ is the opposite of this as we have already discussed, we found that the remaining four source parameters $\left\{\theta, \phi, \mathcal{M}, \omega_0 \right\}$ could benefit from both improvements to timing and pulsar distance uncertainties.


    \subsection{Source localization}\label{subsec: Source Localization}
    
    As we investigated in the previous section, high-precision measurements of pulsar distances (to within $\sigma_L \sim \Delta L_1$) allow for the direct measurement of the source distance $R$ purely through Fresnel corrections.  In this section, we show that this level of precision can also improve localization of the source on the sky.  As we will explain, this improved sky localization mostly comes from phase and frequency parallax terms which can be exploited when the pulsar distance measurements are well-constrained.  When we combine the sky localization with the recovered distance information, there exists potential for future PTAs to pinpoint the galaxy source of the gravitational waves, which could be of great benefit to astronomers seeking electromagnetic counterparts to these coalescing binary sources.
    
    To measure the solid angle on the sky that our uncertainty in the measurement of the source angles $\theta$ and $\phi$ sweeps out, we first compute the confidence ellipse area for those two parameters $\Delta A = \pi \chi^2 \sigma_{\theta^{'}} \sigma_{\phi^{'}}$ as outlined in~\citet{fisher_confidence_ellipses}\footnote{Note that in~\citet{fisher_confidence_ellipses}, equation 7 has a typo in that it is missing the factor of $\chi^2$, which we have corrected here in equation~\ref{eqn: sky localization}.}.  Here, $\sigma_{\theta^{'}}$ and $\sigma_{\phi^{'}}$ are the measured uncertainties along the principle axes of the ellipse itself, and this quantity represents an area in the $\theta$-$\phi$ parameter space.  On the sky, we can define a small solid angle as $\Delta \Omega \approx \sin(\theta) \Delta\theta \Delta\phi$.  So to measure a small confidence ellipse on the sky we connect these two quantities by multiplying $\Delta A$ by $\sin(\theta)$ to get:
    \begin{equation}
        \Delta \Omega \approx \pi \chi^2 \sin(\theta) \sigma_{\theta^{'}} \sigma_{\phi^{'}} = \pi \chi^2 \sin(\theta) \sigma_\theta \sigma_\phi \sqrt{1 - \rho^2} .
    \label{eqn: sky localization}
    \end{equation}
    The uncertainties $\sigma_\theta$ and $\sigma_\phi$ and the correlation coefficient $\rho$ are all measured from the inverse of the Fisher matrix.  In all of our results here, we set $\chi^2 = 2.279$, which gives the approximate $68$ per cent likelihood area for our source.  In a similar calculation for small-volume $\Delta V \approx \frac{4\pi}{3} R^2 \sin(\theta) \Delta R \Delta \theta \Delta \phi$, we compute the volume uncertainty region for our source as:
     \begin{equation}
         \Delta V \approx \frac{4\pi}{3}\chi^3 R^2 \sin(\theta) \sigma_{R^{'}} \sigma_{\theta^{'}} \sigma_{\phi^{'}},
     \label{eqn: volume localization}
     \end{equation}
     where again, primes on the uncertainty variables denote that they are the uncertainties measured along the principle axes of the ellipsoid.  In 3D space, the approximate $68$ per cent likelihood volume of our source corresponds to a $\chi^2 = 3.5059$, which is what we used for all of our results shown here.  The values of the uncorrelated uncertainties along the principle axes $\sigma_{R^{'}}$, $\sigma_{\theta^{'}}$, and $\sigma_{\phi^{'}}$ were found by applying SVD to the 3D sub-matrix of our inverted Fisher matrix (for parameters $R$, $\theta$, and $\phi$).  This mathematically decomposes the matrix into three matrices, one of which is a rectangular diagonal matrix that contains the singular values, that is the uncorrelated uncertainties along the principle axes.
    
    In all of our timing regimes, part of the measurement of the sky angles $\theta$ and $\phi$ comes from the antenna patterns contained in the amplitudes of the Earth and pulsars terms of the timing residuals (see equation~\ref{eqn: antenna patterns}).  Additionally, all timing regimes also have the `phase parallax' term $\left(1-\hat{r}\cdot\hat{p}\right) \frac{L}{c}$, which contains the sky angles, and appears in the phase $\Theta(t)$ of the pulsar terms of each model (see equations~\ref{eqn: Res(t) pw mono phase E and P}, \ref{eqn: Res(t) pw freq evo phase E and P}, \ref{eqn: Res(t) fresnel mono phase E and P}, and~\ref{eqn: Res(t) fresnel freq evo phase E and P}).  In the frequency evolution regimes IB and IIB, this becomes a `phase-frequency parallax' term, because it does appear not only in the pulsar-term phase $\Theta(t)$ but also in the pulsar-term frequency $\omega(t)$. Lastly, the Fresnel regimes add an additional small-order correction $\frac{1}{2}\left(1-\left(\hat{r}\cdot\hat{p}\right)^2\right)\frac{L}{c}\frac{L}{R}$ to these phase/phase-frequency parallax terms.
    
    With this in mind, we looked at examples of improvements to three PTA qualities that can improve sky localization:  timing precision $\sigma$, pulsar distance precision $\sigma_L$, and PTA distance scale factor.  Timing precision helps to measure the sky angles from the overall amplitude of the timing residual.  Pulsar distance precision helps measure the sky angles from the phase/phase-frequency parallax terms due to the $\frac{L}{c}$ factor.  However, this requires high-precision measurements due to the pulsar distance wrapping problem.  It is partially for this reason that studies such as~\citet{NG_11yr_cw} choose to group these parallax terms into separate phase parameters for every individual pulsar in their models (which is a similar but alternative approach to introducing the pulsar \textit{distances} as free parameters in the model, as discussed in Sections~\ref{sec:fisher matrix analysis} and~\ref{sec:L-wrapping problem}).  This approach sacrifices the additional sky localization information contained in these parallax terms to help avoid the wrapping problem.  Finally, increasing the PTA scale factor can help to improve the sky angle measurements by amplifying the chirping effects (due to the greater disparity between the Earth/pulsar frequencies) and boosting the size of the Fresnel corrections.
    
\begin{figure}
        \centering
        \includegraphics[width=0.9\linewidth]{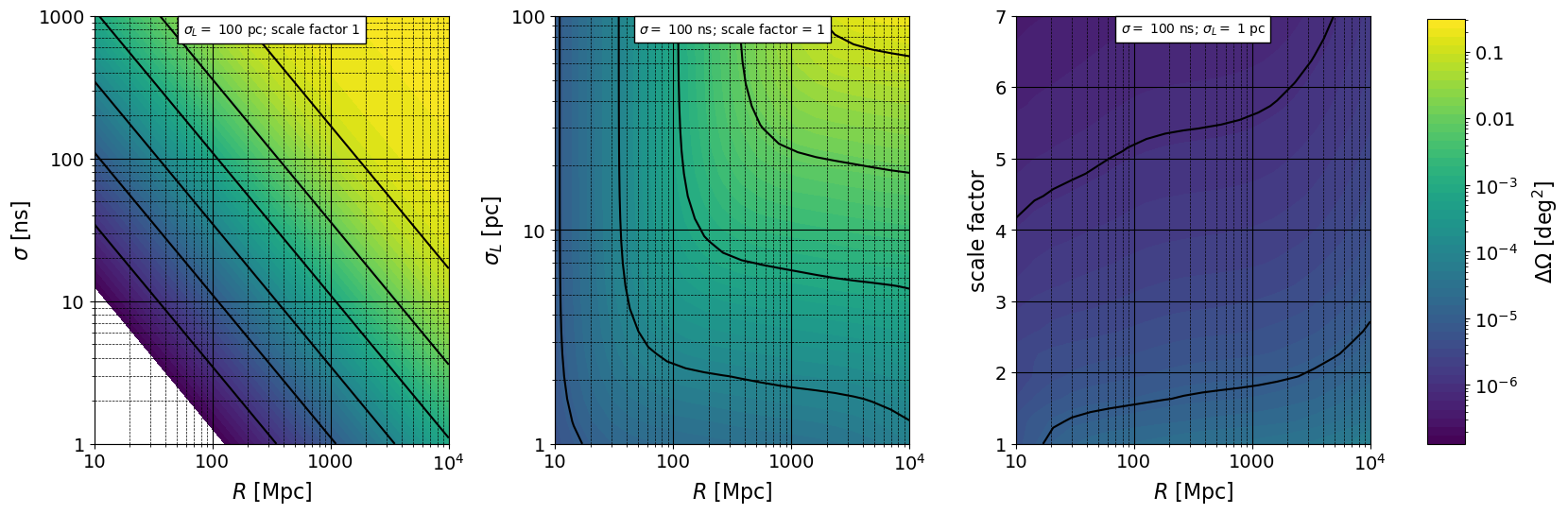}
    \caption{Various effects of PTA improvements on source sky localization, shown here for the example of Source 1.  The bottom left corner of the first panel is excluded because our SVD matrix inversion failed our condition number requirement (see Section~\ref{sec:fisher matrix analysis}); therefore, the Fisher matrix for these cases could not be stability inverted.  Nevertheless, we can clearly see that the trend is that the overall sky accuracy improves as the source gets closer to us and the value of $\sigma$ decreases, as we would expect.  Compared to the other PTA parameters, timing precision improvements behave as a univariate function of $\sigma R$.  Furthermore, timing precision controls the largest range of magnitude differences in $\Delta \Omega$ as a function of source distance, followed by pulsar distance precision, and then finally PTA scale size.  These results were computed using the IIB model (Section~\ref{subsec: fresnel, freq evolution (conjecture)}).}
        \label{fig: sky localization - comparison (contours)}
    \end{figure}

    Examples of improvements in these three PTA qualities as applied to Source 1 are shown in Figure~\ref{fig: sky localization - comparison (contours)}.  It is very difficult to disentangle the many different factors that are all competing to change the measurement of the source's sky localization (including the intrinsic source parameters that we discuss below).  Therefore, we emphasize here that Figure~\ref{fig: sky localization - comparison (contours)} is meant to serve only as one specific illustration of the general comments that we have made so far, for the PTA setups indicated in each panel, and for one particular source.
    
    We find that unlike pulsar distance precision and PTA scale factor, timing precision behaves as a univariate function of $\sigma R$.  In the first panel of Figure~\ref{fig: sky localization - comparison (contours)}, we expect that most of the knowledge of our sky angles is coming from the overall amplitude of the timing residual and not from the phase-frequency parallax or Fresnel corrections (due to the large $\sigma_L$ and standard scale factor).  The second and third panels show the additional benefits that can be gained when probing the information in the phase-frequency parallax and Fresnel corrections.  Recall that $\lambda_\mathrm{gw} \sim 1$ pc for Source 1.  In the middle panel, the Fresnel corrections are not likely adding a great deal of information towards the sky angles themselves since $\sigma_L > \lambda_\mathrm{gw}$ for most of these cases.  However, we still see that information can be gained through the phase-frequency parallax as the precision of the distance measurements to our pulsars improves.  As $\sigma_L$ decreases, we see for a given $\Delta \Omega$ resolution that there are turnover points where even small improvements to the pulsar distance measurements can allow the same source to be localized with that precision at much greater distances.  Finally, with pulsar distance measurements on the order of the source's gravitational wavelength, the right-hand panel shows that increasing the baseline distance between pulsars further improves our ability to measure the source's sky angles.  Overall, we see that timing precision controls the largest range of magnitude differences in $\Delta \Omega$ as a function of source distance, followed by pulsar distance precision, then finally PTA scale size.
    
\begin{figure}
        \centering
        \includegraphics[width=0.45\linewidth]{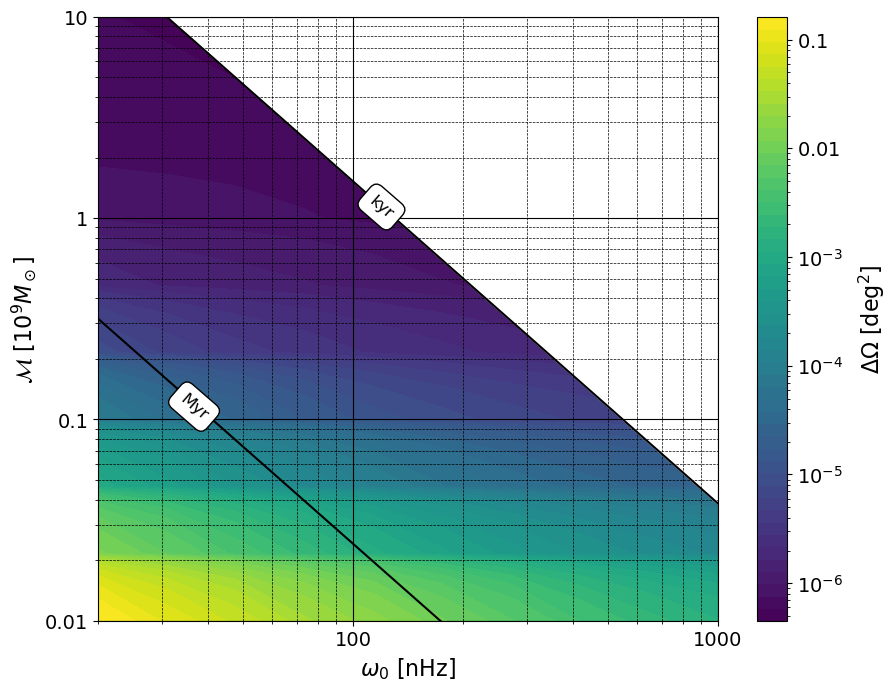}
        \caption{Source sky localization, as a function of source frequency and chirp mass.  These are for sources at $R=100$ Mpc, with the same angular parameters as our fiducial sources (indicated in Table~\ref{tab: source parameters reference}), timing uncertainty $\sigma = 100$ ns, pulsar distance uncertainty $\sigma_L = 1$ pc, and a PTA scale factor of $7$.  Contours of coalescence time $\Delta \tau_c$ (equation~\ref{eqn: coalescence time/angle}) are indicated, with a cut along $\Delta \tau_c = 1$ kyr due to assumption~\ref{as: t_obs << tau_c}.  Bias leans towards lower frequencies, where using a fixed $\sigma_L$ value will benefit sources with larger wrapping cycles as compared to smaller ones (see Section~\ref{subsubsec: distant sources}).  These results were computed using the IIB model (Section~\ref{subsec: fresnel, freq evolution (conjecture)}).  When we re-calculated this plot using the IB model (Section~\ref{subsec: plane-wave, freq evolution}) and IA model (Section~\ref{subsec: plane-wave, monochromatic}), we found that the relative difference in $\Delta \Omega$ between these two models and the original fully general IIB model was about $0.1$ and $0.7$, respectively.}
        \label{fig: sky localization - M-omega0}
    \end{figure}

    The source's intrinsic parameters $\mathcal{M}$ and $\omega_0$ also affect localization.  Figure~\ref{fig: sky localization - M-omega0} shows that for the same distance (here $R=100$ Mpc) and PTA setup, different sources are localized to $\sim\mathcal{O}\left(0.1 \ \mathrm{deg}^2 \ - \ 10 \ \mathrm{arcsec}^2\right)$.  Notice that this PTA setup matches that in the right-hand panel of Figure~\ref{fig: s1 scaled PTA / M-omega0 space sf7 PTA}, so we know that the Fresnel corrections are being probed for the sources where $\mathrm{CV}_R < 1$, since the distance parameter $R$ is measured entirely from these corrections in our model.

    In general, we conclude that the Fresnel corrections themselves do not significantly improve sky localization of a source.  Figure~\ref{fig: sky localization - M-omega0} shows $\Delta \Omega$ predicted from the Fisher matrix analysis using the fully generalized IIB timing residual model (Section~\ref{subsec: fresnel, freq evolution (conjecture)}).  We repeated the same calculations of the sky angles for these sources using the IB model, which removes the Fresnel corrections (Section~\ref{subsec: plane-wave, freq evolution}), and the IA model, which removes both the Fresnel corrections and the frequency evolution effects (Section~\ref{subsec: plane-wave, monochromatic}), in order to compare the predictions between models.  It is important to note that this is not directly comparable, because fundamentally these three models will compute different timing residuals for the same source (see again Figure~\ref{fig: model-regime comparison} - the level of difference will depend on what regime the source is actually in).  And the Fisher matrix analysis asks only how well can we recover the parameters \textit{from a given model}.  Nevertheless, comparing the results of these sets of calculations still gives us an idea of how different the sky localization is with vs. without the Fresnel corrections.  What we found was that the magnitude of the relative difference between $\Delta \Omega$ computed using the models IIB vs. IB and IIB vs. IA was at most about $0.1$ and $0.7$ for what is shown in Figure~\ref{fig: sky localization - M-omega0}, respectively.  This suggests that we lose some sky localization precision when we leave out Fresnel corrections, but we lose more precision when we do not account for frequency evolution.  However, even with these relative differences, we still find that in the IB and IA models (which do not include Fresnel corrections), sky localization ranges from $\sim\mathcal{O}\left(0.1 \ \mathrm{deg}^2 \ - \ 10 \ \mathrm{arcsec}^2\right)$ depending on the intrinsic source parameters.  This along with the earlier observations and statements about Figure~\ref{fig: sky localization - comparison (contours)} is what leads us to believe that sky localization itself is not greatly improved by the inclusion of Fresnel effects.    

    In their separate studies, \citet{CC_main_paper} reported that sky localization could be measured with $\sim\mathcal{O}\left(1-10 \ \mathrm{deg}^2\right)$ while~\citet{DF_main_paper} reported that sky localization could be measured with $\sim\mathcal{O}\left(100 \ \mathrm{arcsec}^2\right)$, which is a significant improvement.  The results and discussion here from our study help to explain why.  These studies use different physical models (with and without Fresnel corrections) and different example sources and PTAs and simulate very different pulsar distance uncertainties.  \citeauthor{CC_main_paper} used values of $\sigma_L \sim \mathcal{O}\left(1-100 \ \mathrm{pc}\right)$, while~\citeauthor{DF_main_paper} used $\sigma_L \sim \mathcal{O}\left(0.001 - 0.01 \ \mathrm{pc}\right)$.  Here we see in Figure~\ref{fig: sky localization - M-omega0} approximately five orders of magnitude difference in sky localization depending on the type of source, and in the middle panel of Figure~\ref{fig: sky localization - comparison (contours)} another five orders of magnitude difference between $\sigma_L = 1 \ - \ 100$ pc.  Therefore, we easily found scenarios where $\Delta \Omega$ ranged from $\mathrm{arcsec}^2$ to $>10\ \mathrm{deg}^2$ precision, simply by accounting for these different effects, PTA qualities, and sources.
    
    While Fresnel corrections in our models may not add much precision to source sky localization, we can combine the distance measured from Fresnel corrections with this sky location to pinpoint our source to within an uncertainty volume of space. An example for Source 1 is shown in Figure~\ref{fig: vol localization - scale factor / S1 localization - vol at 100 Mpc, sky at 1 Gpc}.  The results have only been given where all points have a corresponding $\mathrm{CV} < 1$ for the parameters $R$, $\theta$, and $\phi$.  Since the source distance $R$ is the hardest parameter to measure in general, our ability to localize the source to within some uncertainty volume in space is most directly determined by how well we can measure that parameter.  Some parts of the sky are less sensitive than others, and even if we lose the sensitivity within our PTA required to measure the volume localization $\Delta V$, we can still typically measure a sky location $\Delta \Omega$ with striking precision.  The right-hand panel of Figure~\ref{fig: vol localization - scale factor / S1 localization - vol at 100 Mpc, sky at 1 Gpc} shows that even at $1$ Gpc, Source 1 was measured here with sub-square arcminute precision no matter where it was located on the sky.
    \begin{figure}
        \centering
          \begin{subfigure}[t]{0.48\linewidth}
          \centering
            \includegraphics[width=1\linewidth]{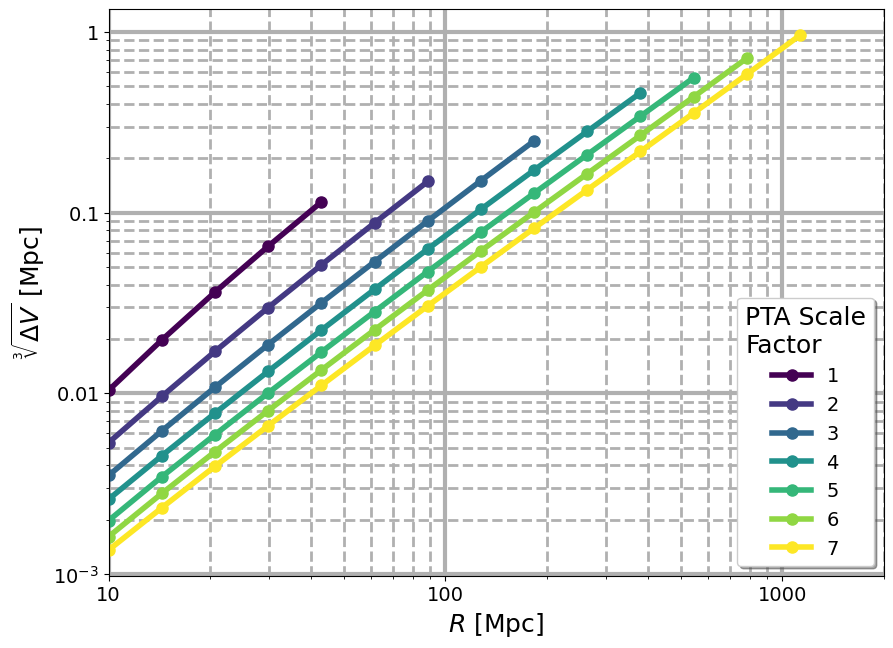}
          \end{subfigure}
          \hfill
          \begin{subfigure}[t]{0.38\linewidth}
          \centering
          \includegraphics[width=1\linewidth]{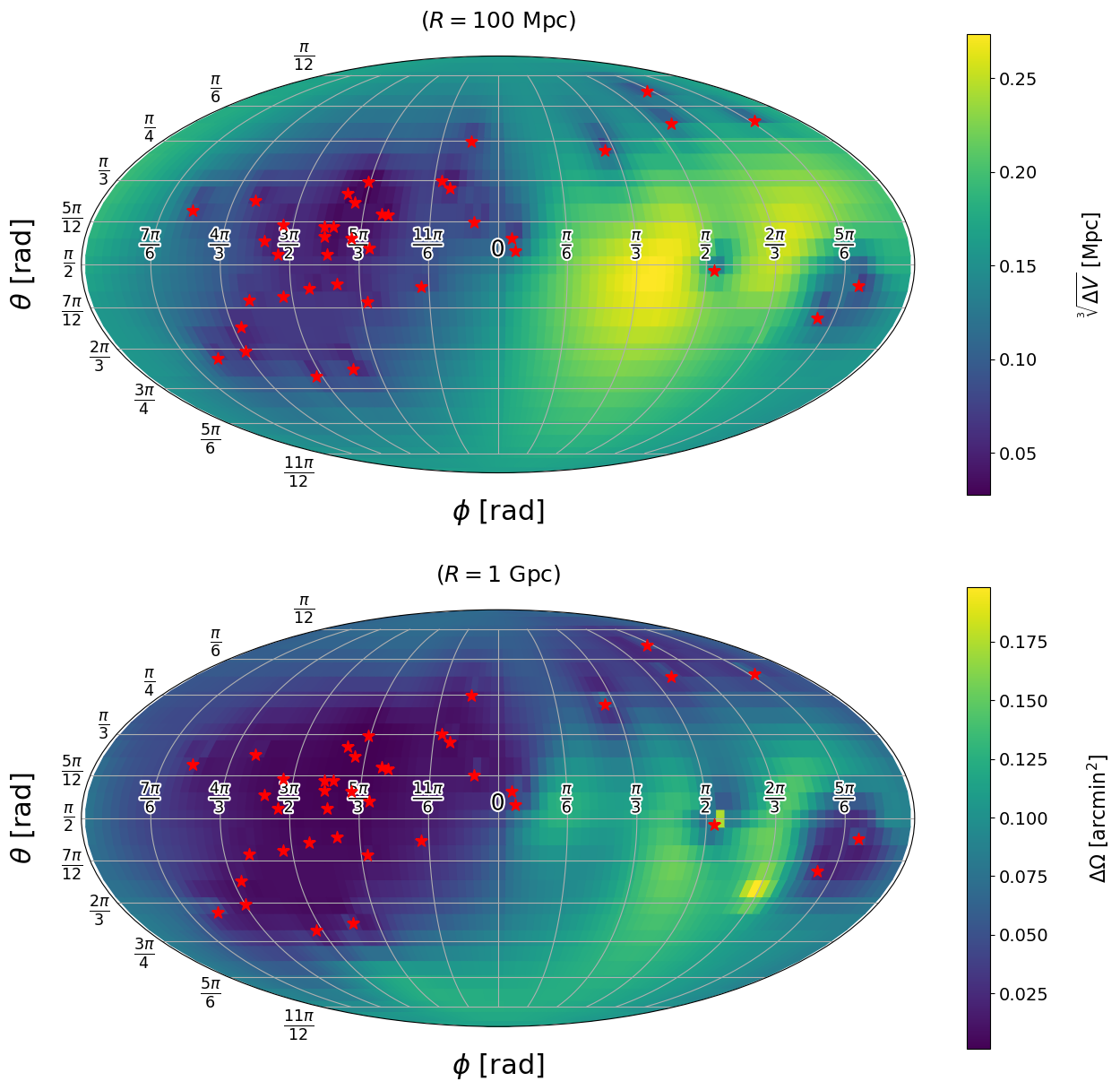}
          \end{subfigure}
    \caption{Source sky and volume and localization calculated for Source 1, with timing uncertainty $\sigma = 100$ ns, and pulsar distance uncertainty $\sigma_L = 1 \ \mathrm{pc}\ \sim \lambda_\mathrm{gw}$.  The areas and volumes represent the $68$ per cent likelihood regions of our source.  These results were computed using the IIB model (Section~\ref{subsec: fresnel, freq evolution (conjecture)}).  \textbf{(Right)}~The top panel shows volume localization of the source at $R=100$ Mpc (therefore, $A_{E,\mathrm{res}}=2 \ \mu$s).  In the bottom panel the source is at $R=1$ Gpc (therefore, $A_{E,\mathrm{res}}=207$ ns), and although we can no longer measure its distance with this PTA (and hence calculate the volume localization), we can still measure the sky angles from the chirping in the signal to localize it on the sky.  In both the top and bottom panels, the PTA has a scale factor of $5$.}
    \label{fig: vol localization - scale factor / S1 localization - vol at 100 Mpc, sky at 1 Gpc}
    \end{figure}


    \subsection{Measuring chirp mass from a monochromatic source}\label{subsec: Measuring Chirp Mass from a Monochromatic Source}
    
    Another important novelty of the Fresnel regime is that it allows a measurement of the source's chirp mass even if the source is producing monochromatic gravitational waves.  For a monochromatic source, Fresnel corrections break the $R$-$\mathcal{M}$ degeneracy in the IIA regime (see Section~\ref{subsec: fresnel, mono}).  As long as we can measure $A_{E,\mathrm{res}}$, $R$, and $\omega_0$, then from equation~\ref{eqn: earth term amplitude}, we can infer the the chirp mass.  As pointed out in Section~\ref{sec:results}, within the limitations of this study, low-frequency sources below $\omega_0 < 10$ nHz require a greater analysis of the impact of other timing residual sources before confidently measuring the gravitational wave parameters and their uncertainties.  Therefore, what we show here is simply meant to be a proof of concept.
    
    For this, we use the fiducial Source 2 in Table~\ref{tab: source parameters reference}.  An ideal monochromatic source would have infinite $\Delta \tau_c$, so we chose a low chirp mass source on the edge of our SNR limitation.  At $R=100$ Mpc, the amplitude of such a source would be very small, so we also simulate a future timing experiment capable of $\sigma \sim 1$ ns timing uncertainties.  In a future study, it would be interesting to simulate the measurement of a higher chirp mass system with an orbital frequency $\omega_0 \sim 1$ nHz, as this type of source would produce a stronger amplitude.
    
\begin{figure}
        \centering
          \begin{subfigure}[t]{0.35\linewidth}
          \centering
          \includegraphics[width=1\linewidth]{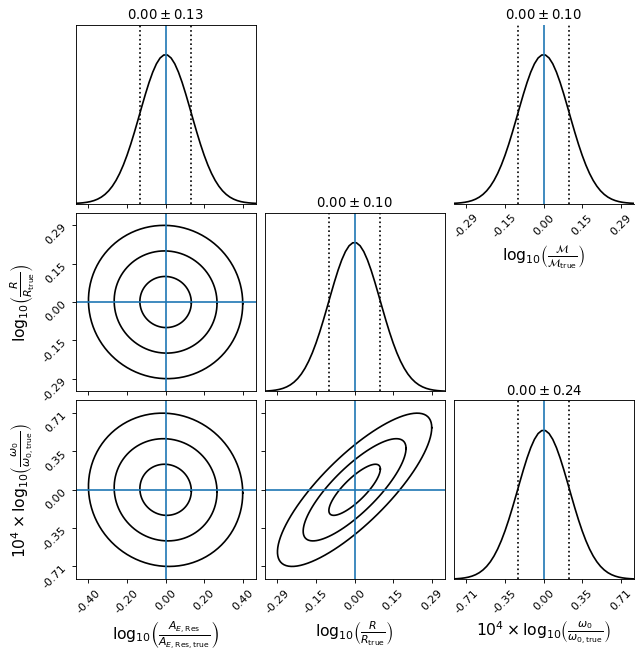}
          \end{subfigure}
          \hfill
          \begin{subfigure}[t]{0.45\linewidth}
          \centering
            \includegraphics[width=1\linewidth]{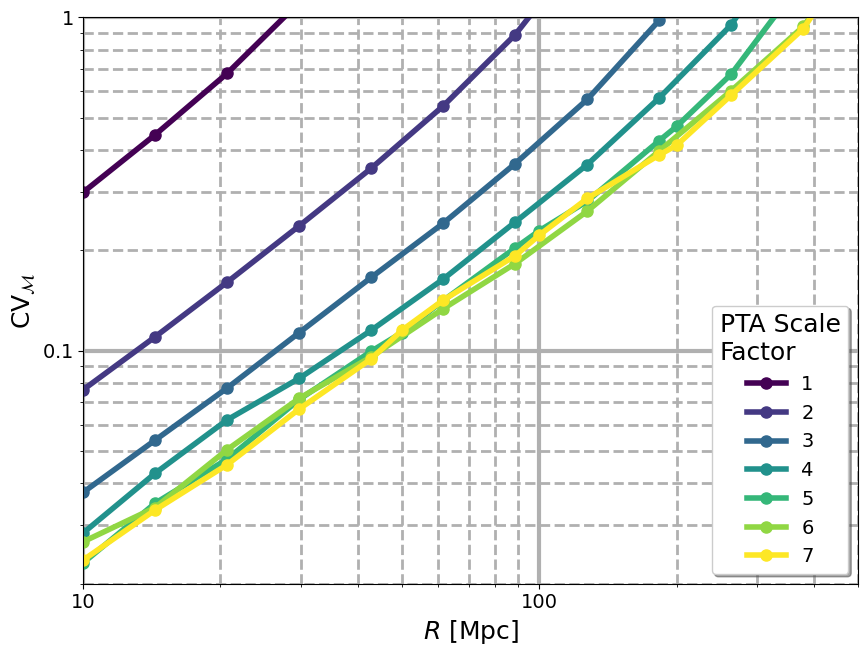}
          \end{subfigure}
    \caption{Error propagation to the measurement of the source's chirp mass $\mathcal{M}$ for a monochromatic source recovered from the measurements of $A_{E,\mathrm{res}}$, $R$, and $\omega_0$, calculated here in both panels for Source 2.  Here, all pulsars have been given timing uncertainty $\sigma = 1$ ns, and pulsar distance uncertainty $\sigma_L = 1.5 \ \mathrm{pc}\ \sim \lambda_\mathrm{gw}$.  Note that in this study, we chose to parametrize the model in the log of the parameters $\left\{A_{E,\mathrm{res}}, \ R, \ \omega_0\right\}$.  These results were computed using the IIA model (Section~\ref{subsubsec: fresnel, mono (Heuristic model)}).  \textbf{(Left)} Here, we show the recovery and correlations of the log of the parameters $\left\{ A_{E,\mathrm{res}}, \ R, \ \omega_0 \right\}$ in the triangle plot, and in the upper right corner, we show the propagated uncertainty on the measurement of the log of $\mathcal{M}$.  Source 2 has been placed at $R=100$ Mpc (therefore, $A_{E,\mathrm{res}}=0.15$ ns), and here the PTA has a scale factor of $5$.  In this particular example, we find that the uncertainty on the chirp mass is dominated by the pulsar distance uncertainty $\sigma_L$, which most strongly affects the recovery of $R$.  The Fisher matrix analysis predicts that this source can be recovered with:
    $\left\{ \mathrm{CV}_{A_{E,\mathrm{res}}}, \ \mathrm{CV}_{R}, \ \mathrm{CV}_\theta, \ \mathrm{CV}_\phi, \ \mathrm{CV}_\iota, \ \mathrm{CV}_\psi, \ \mathrm{CV}_{\theta_0}, \ \mathrm{CV}_{\omega_0} \right\} = \left\{ 0.32, \ 0.23, \ 7\times10^{-6}, \ 4\times10^{-6}, \ 0.48, \	0.58, \ 0.59, 5.4\times10^{-5} \right\}$.  Uncertainty propagation gives an inferred $\mathrm{CV}_{\mathcal{M}} = 0.23$ .  \textbf{(Right)} For small PTA scale factors, recovery of $\mathcal{M}$ is dominated by pulsar distance uncertainty $\sigma_L$.  As the scale factor increases, the larger Fresnel numbers make the measurement of $R$ more precise, and as a result for scale factors above about $5$ the measurements of $\mathcal{M}$ do not improve because they become dominated by timing uncertainty $\sigma$.  Note that a scale factor of $7$ puts the furthest pulsar in our PTA at roughly the distance of the Large Magellanic Cloud.}
    \label{fig: source iia example - M recovery / source iia example - scalefactor plot}
    \end{figure}

    The left-hand panel of Figure~\ref{fig: source iia example - M recovery / source iia example - scalefactor plot} shows an example of the uncertainty propagation to our measurement of the system's chirp mass.  We found that measuring $\mathcal{M}$ was most strongly correlated to measuring $A_{E,\mathrm{res}}$ and $R$.  Since measuring $A_{E,\mathrm{res}}$ depends on the timing uncertainty $\sigma$, and measuring $R$ depends on the pulsar distance uncertainty $\sigma_L$, this means that different experimental setups may depend more on timing or parallax for the recovery of the system chirp mass.  As an example, this can be seen in the right-hand panel of Figure~\ref{fig: source iia example - M recovery / source iia example - scalefactor plot}.  Initially, increasing the PTA scale factor notably improves the recovery of $\mathcal{M}$, since the higher Fresnel numbers and improved parallax measurements allow for better recovery of $R$ (similar to what was shown earlier in the left-hand panel of Figure~\ref{fig: s1 scaled PTA / M-omega0 space sf7 PTA}).  However, around a scale factor of $5$, we reach the threshold where timing precision dominates over parallax in our ability to recover $\mathcal{M}$.  Moreover, for this source, we found that increasing $\sigma > 1$ ns more dramatically made the recovery of $\mathcal{M}$ dependent on the recovery of $A_{E,\mathrm{res}}$.

    \section{Conclusions}\label{sec: conclusions}

In this work, we motivate the importance of Fresnel corrections in the effects of gravitational waves on pulsar timing residual models and develop what we call the `Fresnel' timing regime, separate from the previously established plane-wave regimes, which are currently used in pulsar timing searches for continuous waves~\citep{PPTA_cw_2014,EPTA_cw_2016,NG_11yr_cw}.  We derive the Fresnel monochromatic model analytically, study it asymptotically, and provide a more simple and heuristically motivated model that is easier to implement in timing simulations and analyses.  With this, we build a well-motivated conjecture of the Fresnel frequency evolution timing residual model, which fully generalizes the previously studied plane-wave frequency evolution models with Fresnel order corrections to the phase and frequency of the timing residual.  Surrounding this, we also offer a framework for understanding the relevant limits of each model, namely by considering the coalescence time of the source and the size of its Fresnel number with the pulsars in the timing array.

Then, we perform a Fisher matrix analysis for parameter estimation within these Fresnel regimes, using a representative NANOGrav-related PTA as the base for the experimental design within our simulations.  This was meant to build upon and bridge the gap between the studies by~\citet{CC_main_paper} and~\citet{DF_main_paper}.  The main challenge in measuring the source distance purely from Fresnel corrections is the pulsar distance wrapping problem.  We offer a new look at this problem and show how well measured the pulsar distances will need to be in order to accurately recover the source distance from our searches.  In general, distance constraints are needed on the order of the wrapping cycle on the most distant pulsars in the PTA but not on the nearby pulsars whose Fresnel corrections are negligible.

One way to increase the Fresnel corrections in the timing residuals is by including pulsars at farther distances, since the Fresnel number scales as $F \propto L^2$ (see equation~\ref{eqn: Fresnel number}).  Intrinsic source properties $\mathcal{M}$ and $\omega_0$ also play an important role as they control both the amount of frequency evolution in the residuals and the characteristic wrapping cycle.  In our simulations, we find that future PTAs with distant pulsars across the Milky Way Galaxy whose distances are constrained to the order of the wrapping cycle can measure the source distance with a $\mathrm{CV}_R \la 0.1$ for sources at out to $\mathcal{O}(100 \ \text{Gpc})$ distances.

High-precision pulsar measurements not only allow us to probe the Fresnel corrections but also help to localize the source through the combined phase-frequency parallax of all of the pulsars in the PTA.  The previous studies of source localization by \citeauthor{DF_main_paper} and \citeauthor{CC_main_paper} had shown many orders of magnitude difference in the sky angle ranging from $\mathcal{O}(100 \ \text{arcsec}^2 \ - \ 10 \text{deg}^2)$.  In this work, we explain that this difference is due to many factors including the timing accuracy $\sigma$, the pulsar distance accuracy $\sigma_L$, the PTA scale factor, and the intrinsic source parameters.  We also show how knowledge of the source distance through Fresnel corrections can be combined with the sky angle measurements in order to localize the source to a volume of space.  Even for sources near $\sim 1$ Gpc in distance could be localized to within a volume $\Delta V < 1 \ \text{Mpc}^3$.  This would allow us to localize the galaxy source of the gravitational waves, which could then be targeted with an in-depth multimessenger follow-up, such as what was done recently in \citet{multimessenger_caitlin_2020}.

A novelty of the Fresnel formalism is that it offers a way of measuring the chirp mass of a monochromatic source, which is something that cannot be done using the plane-wave formalism.  As long as the amplitude, distance, and frequency parameters can be measured, the chirp mass can be calculated.  Furthermore, the Fresnel formalism opens a path for future cosmological studies using PTAs.  As discussed recently in~\citet{DOrazio_main_paper}, when the Fresnel models are generalized to a cosmologically expanding universe, the distance parameter $R$ that comes from the Fresnel corrections becomes the comoving distance $D_c$.  If the source is also significantly chirping, then the system's observed (redshifted) chirp mass can be measured.  Combining this with the measured observed frequency and amplitude parameters produces a measurement of the system's luminosity distance $D_L$, which is related to the comoving distance through $D_L = (1 + z)D_c$.  Therefore, in the full Fresnel frequency evolution regime IIB, frequency evolution effects allow for the measurement of the source $D_L$, and Fresnel corrections allow for the measurement of the source $D_c$.  If both of these distances are recovered from the timing residual model, then a measurement of $H_0$ can be obtained (as shown in \citeauthor{DOrazio_main_paper}).  Or alternatively, if only one of these two distances can be measured accurately, but localization of the gravitational wave source and multimessenger counterparts identify the host galaxy, then the source host galaxy's redshift can be measured.  Combining the redshift with either of the measured source distances once again provides a measurement of $H_0$.

While we acknowledge that many of our `best case' scenarios required simulating idealized PTAs well beyond our current experimental capabilities, timing and pulsar measurements will continue to improve~\citep{Lam_2018}.  The Five hundred meter Aperture Spherical Telescope is now operational and offers exciting prospects for discovering many new pulsars~\citep{FAST_smits2009}.  The MeerKAT and MeerTime projects could provide high-precision timing of pulsars with absolute timing errors on the order of 1 ns~\citep{MeerKAT_bailes2020}, and the future Square Kilometer Array is going to provide even greater sensitivity and will greatly increase the number of pulsars that we can time~\citep{SKA_janssen2015}.  Furthermore, \citet{GWastro_Lee2011} and~\citet{Ldist_smitts2011} discuss how pulsar distance measurements can be improved through timing parallax, which will improve with these more sensitive future experiments and could help push the pulsar distance precision closer to what is needed for this work.  In general, this suggests the potential for future studies that could be made possible from the Fresnel timing regimes.  For example, seeing the direct impact that PTA size has on parameter estimation hopefully offers strong motivation to actively search for and include more distant pulsars with higher precision distance measurements in our current PTAs.  \citet{FAST_smits2009} even discusses the possibility of detecting extragalactic pulsars.  While it is unlikely that such pulsars would have the precision constraints necessary for the measurements proposed in this work, it still points to new frontiers in the future of PTA science, which could make this work possible.

Finally, we reiterate here that in this work, we have explicitly assumed a flat static universe for the sake of simplicity and to help isolate the measurement of the distance parameter $R$ [see assumption~\ref{as:empty}].  More generally, for an expanding universe, the fully generalized Fresnel frequency evolution model contains both a luminosity and a comoving distance parameter.  We will explore this generalized model in greater detail in a follow-up study to this paper and the intriguing cosmological implications that the Fresnel regime may offer to future PTAs.

\section*{Acknowledgements}

We thank Luke Kelley for helpful comments during the preparation of this paper.  This work was supported by the National Science Foundation (NSF) through the NANOGrav collaboration Physics Frontier Center award NSF PHY-1430284, as well as the awards NSF PHY-1607585 and NSF PHY-1912649.  This work also received support from the Wisconsin Space Grant Consortium Graduate and Professional Research Fellowship Program under the NASA Training Grant \#NNX15AJ12H, as well as from the University of Wisconsin-Milwaukee, for which we are grateful for access to their computational resources supported by NSF PHY-1626190.

\section*{Data Availability}
Calculations in this paper were performed using code developed by the authors, and the data are available on reasonable request to the authors.




\bibliographystyle{mnras}
\bibliography{references}



\pagebreak
\section*{Supporting Information}
Supplementary data are available at MNRAS online.

\noindent Please note: Oxford University Press is not responsible for the content or functionality of any supporting materials supplied by the authors.
Any queries (other than missing material) should be directed to the corresponding author for the article.

\noindent\textbf{Figure S1} The fractional difference in the gravitational wave timing residual SNR defined in equation 1, with versus without a quadratic polynomial best fit subtracted from the data, as a function of the source orbital frequency $\omega_0$ and chirp mass $\mathcal{M}$.

\noindent\textbf{Figure S2} Fisher matrix measurements of the parameters for Source 1, which has been placed at $R=500$ Mpc (therefore, $A_{E,\mathrm{res}}=414$ ns).




\appendix

    \section{Deriving the Gravitational Wave Timing Residual}\label{app: deriving the timing residual}

In this section, we give a brief overview of the calculations that go into modelling the gravitational wave-induced timing residual for a pulsar timing (or one-way Doppler tracking) experiment.  There exist various approaches to the derivation of the residual in the literature~\citep{Finn_2009,creighton_anderson_2011}.  The approach we take is to study the path of a photon by integrating the space-time metric, as shown in~\citet{maggiore_2018}.  \citet{Finn_2009} notes that, under certain circumstances, the method used here will not be valid; however, as we assume a flat Minkowski background (see below) and express the metric in the transverse-traceless gauge, the conditions for validity required in~\citet{Finn_2009} are satisfied.  Below we indicate the assumptions that go into this work and our derivations.


    \subsection{Assumptions}\label{app: assumptions}
    
    \textit{Einstein Field Equations and Source Binary}
    \begin{enumerate}
        \item Cosmologically static and flat universe (hence the background metric is flat Minkowski, ${\eta_{\mu \nu} = \mathrm{diag}(-c^2, \ 1, \ 1, \ 1)}$). \label{as:empty}
        \item Weak field limit (gravitational waves are a metric perturbation on top of the Minkowski background). \label{as:weak}
        \item Small source compared to the distance to the observer and the wavelength of the wave. \label{as:small}
        \item Slow-moving source (non-relativistic, no post-Newtonian analysis required). \label{as:slow}
        \item Transverse-traceless gauge. \label{as:TT-gauge}
        \item The binary source is circular. \label{as:circular}
        \item The source is located at a fixed angular sky position and distance from the Earth.
        \item Far-field approximation of the metric perturbation \textit{amplitude} - the binary source is sufficiently far from the field point of interest that $|\vec{x}_\mathrm{field} - \vec{x}_\mathrm{source}| \approx R$ in the amplitude of the metric perturbation.  This assumption will not be made when evaluating the retarded time.
    \end{enumerate}
    \textit{Earth and Pulsar}
    \begin{enumerate}\setcounter{enumi}{9}
        \item Earth is at the centre of our coordinate system.
        \item The pulsar is located at a fixed angular sky position and distance from the Earth.
        \item The antenna patterns are assumed to remain constant over the Earth-pulsar baseline.
        \item The pulsar's own rotation period $T$ time-scale must be appropriately small in comparison to the orbital period and coalescence time of the gravitational wave source.  Specifically, we have two case requirements depending on the gravitational wave source regime: \label{as: pulsar period timescale}
        \begin{itemize}
            \item \underline{Monochromatic Source}:  $\omega_0 T \sim \frac{T}{T_\mathrm{orbit}} \ll 1$  
            \item \underline{Frequency Evolving Source}:  $\frac{T}{\Delta \tau_c} \ll 1$
        \end{itemize}
        Our pulsars are millisecond pulsars~\citep{NG_11yr_data} and the orbital period of our sources of interest is $\mathcal{O}\left(\text{weeks} \ - \ \text{decades}\right)$, so this assumption is safe.
    \end{enumerate}
    \textit{Frequency Evolution Formalism}
    \begin{enumerate}\setcounter{enumi}{13}
        \item The observation time scale is much, much less than the time to coalescence (measured from the fiducial time): $t_\mathrm{obs} \ll \Delta \tau_c$ \label{as: t_obs << tau_c}
    \end{enumerate}
    \textit{Fresnel Formalism}
    \begin{enumerate}\setcounter{enumi}{14}
        \item When deriving the Fresnel formalism, we assume that we are not working in any of the plane-wave limits, that is when $F=0$ or $\ \hat{r}\cdot\hat{p} = \pm 1$.
        \item Corrections in the geometrical orientation components that affect the amplitude of the timing residual, namely the antenna patterns, which are due to the curvature of a gravitational wave are negligible.  Only corrections that directly affect the phase and frequency of these quantities are significant.
    \end{enumerate}


    \subsection{The Gravitational Wave Timing Residual}
    We solve for the model of an induced gravitational wave timing residual on a pulsar by first considering the time of arrival of photons in successive pulsar periods.  Photons will travel the radial null geodesic between the pulsar and the Earth.  We start with the space-time metric:
    \begin{equation}
        0 = -(c dt)^2 + \left[ \eta_{ij} + h^{TT\hat{r}}_{ij} \right] \hat{p}^i\hat{p}^j dr^2  ,
    \label{eqn: radial null spacetime interval}
    \end{equation}
    where $h^{TT\hat{r}}_{ij} = h^{TT\hat{r}}_{ij}\big(t_\mathrm{ret}(t,\vec{x})\big)$ is evaluated at the retarded time of the gravitational wave.  The superscript $TT\hat{r}$ denotes that we are in the transverse-traceless gauge along some axis $\hat{r}$, which is the direction to our source, and $\hat{p}$ is the direction to our pulsar.  The photon's path to zeroth order in the metric perturbation is $\vec{x}_0(t) \approx \left[ct_\mathrm{obs} - ct\right]\hat{p}$, and we integrate along that path between the endpoints ${(t,r) = (t_\mathrm{obs}-\frac{L}{c}, L) \rightarrow (t_\mathrm{obs}, 0)}$.  This gives us an expression for the time the first photons are observed on Earth after they were emitted from the pulsar. The integral is then repeated for the photons one pulsar period $T$ later.  The difference in these two expressions gives us the observed pulsar period on Earth, $T_\mathrm{obs} \approx T + \Delta T$, where:
    \begin{equation}
        \Delta T =\frac{1}{2}\hat{p}^i\hat{p}^j E^{\hat{r}\textsc{A}}_{ij}\int\limits^{t_\mathrm{obs}}_{t_\mathrm{obs}-\frac{L}{c}} \bigg[h_{\textsc{A}}\Big(t_{\mathrm{ret}}\big(t+T, \vec{x}_0(t)\big) \Big) - h_{\textsc{A}}\Big(t_{\mathrm{ret}}\big(t, \vec{x}_0(t)\big) \Big) \bigg]dt .
    \label{eqn: Delta T}
    \end{equation}
    Dividing both sides of this expression by $T$ we can write $\frac{\Delta T}{T}$, which is the gravitational wave-induced fractional shift in the pulsar period.  Next, we take the resulting integrand and recognize that as long as the non-dimensional quantities indicated in assumption~\ref{as: pulsar period timescale} are small, we can approximate that integrand as a derivative.  This allows us to now write:
    \begin{equation}
        \frac{\Delta T}{T}(t_\mathrm{obs}) \approx \frac{1}{2}\hat{p}^i\hat{p}^j E^{\hat{r}\textsc{A}}_{ij} \int\limits^{t_\mathrm{obs}}_{t_\mathrm{obs} - \frac{L}{c}} \frac{\partial h_{\textsc{A}}\big(t_\mathrm{ret}(t,\vec{x})\big)}{\partial t}\Bigg\rvert_{\vec{x}=\vec{x}_{0}(t)} dt .
    \tag{\ref{eqn: Delta T / T} r}
    \end{equation}
    
    For typical pulsars and sources of interest, the period shift is far too small to observe directly.  However, over long periods of time, the period shifts of every pulsar period accumulate and cause the overall time of arrival (TOA) of a pulsar's pulse to drift sinusoidally in time.  This integrated effect is known as the timing residual, and mathematically it is the integral of the fractional period shift over the observation time-scale, which is equal to the difference in the observed TOA and the expected TOA at a given time $t$:
    \begin{equation}
        \mathrm{Res}(t) = \int\frac{\Delta T}{T}\left(t_\mathrm{obs}\right) dt_\mathrm{obs} = \int\frac{T_\mathrm{obs} (t_\mathrm{obs}) - T}{T} dt_\mathrm{obs}  = \ \mathrm{Obs}(t) - \mathrm{Exp}(t) .
    \tag{\ref{eqn:timing residual} r}
    \end{equation}


    \subsection{Fresnel, Monochromatic Regime}
    In order to solve for the residual in the Fresnel regime, we first need to evaluate the retarded time equation~\ref{eqn:retarded time} along the photon's path $\vec{x}_0(t)$:
    \begin{align}
        t_\mathrm{ret}^0 \equiv t_\mathrm{ret}\big(\vec{x}=\vec{x}_0(t)\big) &= t - \frac{R}{c}\sqrt{1 - 2\hat{r}\cdot\hat{p}\left(\frac{ct_\mathrm{obs} -ct}{R}\right) + \left(\frac{ct_\mathrm{obs} -ct}{R}\right)^2} , \nonumber \\
        &= t \ - \frac{R}{c} \ +  \left(\hat{r}\cdot\hat{p}\right)\frac{ct_\mathrm{obs} -ct}{c} \ -  \frac{1}{2}\left(1-\left(\hat{r}\cdot\hat{p}\right)^2\right)\frac{\left(ct_\mathrm{obs} -ct\right)^2}{cR} \ + \ \ldots 
       \label{eqn:tret 0}
    \end{align}
    To derive the plane-wave regime models given in Sections~\ref{subsec: plane-wave, monochromatic} and~\ref{subsec: plane-wave, freq evolution}, we need to keep only the first three terms in this expansion.  However, in order to study the Fresnel regime, we need to keep out to the fourth term in equation~\ref{eqn:tret 0}.   The result can be re-expressed with a change of variables in the form:
    \begin{align}
        t^0_\mathrm{ret} &\approx \left(t_\mathrm{obs}-\frac{R}{c}\right) - \frac{R}{c}\left(1-\hat{r}\cdot\hat{p}\right)u - \frac{1}{2}\frac{R}{c}\left(1-\left(\hat{r}\cdot\hat{p}\right)^2\right)u^2 , \label{eqn:tret 0 fr} \\
        &\mathrm{where}\quad u \equiv \frac{ct_\mathrm{obs}-ct}{R} . \nonumber
    \end{align}
    
    It is helpful to remember the complete functional dependence of the metric perturbation $h_{\textsc{A}} = h_{\textsc{A}}\bigg(h\Big(\omega\big(t_\mathrm{ret}(t,\vec{x})\big)\Big), \Theta\big(t_\mathrm{ret}(t,\vec{x})\big)\bigg)$ when applying the chain rule in the next step and when solving the integral below in equation~\ref{eqn: Delta T / T fr mono integral} (see equation~\ref{eqn: h+x(t) & h(t)}, and remember from equation~\ref{eqn: radial null spacetime interval} that the metric perturbation is evaluated at the retarded time of the gravitational wave).  For the case of a monochromatic gravitational wave, the phase and the frequency are given in equation~\ref{eqn: monochrome model}.  With this, the integrand in equation~\ref{eqn: Delta T / T} can be written as $\frac{\partial h_{\textsc{A}}\big(t, \vec{x} \big)}{\partial t}\Big\rvert_{\vec{x}=\vec{x}_{0}(t)} = \frac{\partial h_{\textsc{A}}}{\partial \Theta}\Big\rvert_{\vec{x}=\vec{x}_{0}(t)}\omega_0$.  Next, we can perform a change of variables in our integral to integrate over $u$ as defined in equation~\ref{eqn:tret 0 fr}.  Putting all of this together allows us to write equation~\ref{eqn: Delta T / T} as:
    \begin{align}
           \frac{\Delta T}{T} &= \frac{1}{2}\hat{p}^i\hat{p}^j E^{\hat{r}\textsc{A}}_{ij} \frac{R}{c} \int\limits^{L/R}_{0} \frac{\partial h_{\textsc{A}}}{\partial \Theta} \Bigg\rvert_{\vec{x}=\vec{x}_{0}(t)} \omega_0 du , \nonumber \\
            &= \hat{p}^i\hat{p}^j E^{\hat{r}\textsc{A}}_{ij}\omega_0\frac{R}{c} h_0 g_\textsc{A} , \label{eqn: Delta T / T fr mono integral}\\
            &\underset{\left(t_0 \ = \ -\frac{R}{c}\right)}{\mathrm{where}}\quad \begin{cases}
            g_{+} &\equiv \int\limits^{L/R}_0 \sin\left(\textsf{A} - \textsf{B}u - \textsf{C}u^2\right) du ,\\
            g_{\times} &\equiv -\int\limits^{L/R}_0 \cos\left(\textsf{A} - \textsf{B}u - \textsf{C}u^2\right) du , \\
            \textsf{A} &\equiv 2\theta_0 + 2\omega_0 t_\mathrm{obs} , \\
            \textsf{B} &\equiv 2\omega_0\frac{R}{c}\left(1-\hat{r}\cdot\hat{p}\right) = \frac{2\textsf{C}}{\left(1+\hat{r}\cdot\hat{p}\right)} , \\
            \textsf{C} &\equiv \omega_0\frac{R}{c}\left(1-\left(\hat{r}\cdot\hat{p}\right)^2\right) = \frac{\left(1+\hat{r}\cdot\hat{p}\right)\textsf{B}}{2} .
        \end{cases}
    \end{align}
    These resulting integrals can be solved by appealing to complex methods and by using the definitions of the Fresnel integrals:
    \begin{align}
        \begin{cases}
        S(\eta) \equiv \int\limits^\eta_0 \sin\left(\frac{\pi}{2} x^2\right) dx \quad \xrightarrow[|\eta| \gg 1]{\text{asymptotic expansion}} \quad \frac{\mathrm{sgn}(\eta)}{2} - \frac{1}{\pi \eta}\cos\left(\frac{\pi}{2}\eta^2\right) , \\
        C(\eta) \equiv \int\limits^\eta_0 \cos\left(\frac{\pi}{2} x^2\right) dx \quad \xrightarrow[|\eta| \gg 1]{\text{asymptotic expansion}} \quad \frac{\mathrm{sgn}(\eta)}{2} + \frac{1}{\pi \eta}\sin\left(\frac{\pi}{2}\eta^2\right) .
        \end{cases}
    \label{equation: Fresnel integrals}
    \end{align}
    The result is:
    \begin{align}
        &\begin{cases}
        g_{+} = \sqrt{\frac{\pi}{2\textsf{C}}} \bigg[ \Big\{C\left(\eta_2\right)-C\left(\eta_1\right)\Big\}\sin(\Phi) - \Big\{S\left(\eta_2\right)-S\left(\eta_1\right)\Big\}\cos(\Phi) \bigg] , \\
        g_{\times} = -\sqrt{\frac{\pi}{2\textsf{C}}} \bigg[ \Big\{C\left(\eta_2\right)-C\left(\eta_1\right)\Big\}\cos(\Phi) + \Big\{\left(\eta_2\right)-S\left(\eta_1\right)\Big\}\sin(\Phi) \bigg] , \\
        \Phi \equiv \textsf{A} + \frac{\textsf{B}^2}{4\textsf{C}} , \\
        \eta_1 \equiv \frac{\textsf{B}}{\sqrt{2\pi\textsf{C}}} , \\
        \eta_2 \equiv \eta_1 \Big[1 + \frac{2\textsf{C}}{B}\left(\frac{L}{R}\right) \Big] ,
        \end{cases}
    \end{align}
    Now we can solve equation~\ref{eqn:timing residual} to get the timing residual.  The only term that depends on $t_\mathrm{obs}$ is $\Phi = \Phi\big(\textsf{A}\left(t_\mathrm{obs}\right)\big)$, which we can use to write $\frac{d\Phi}{dt_\mathrm{obs}}=\frac{d\Phi}{d\textsf{A}}\frac{d\textsf{A}}{dt_\mathrm{obs}} = 2\omega_0$ and perform the following change of variables in the integration:
    \begin{equation}
        \mathrm{Res}(t) = \int \left[\hat{p}^i\hat{p}^j E^{\hat{r}\textsc{A}}_{ij}\omega_0\frac{R}{c} h_0 f_\textsc{A}\right] dt_\mathrm{obs} = \hat{p}^i\hat{p}^j E^{\hat{r}\textsc{A}}_{ij}\omega_0\frac{R}{c} h_0 \int f_\textsc{A} \frac{d\mathrm{\Phi}}{2\omega_0} .
    \end{equation}
    The final result of this integration is given in equation~\ref{eqn: Res(t) fresnel mono}.

    \section{Data Reference Tables}\label{app: data reference}

\vspace{7cm}

\begin{table}
\centering
    \caption{These are the parameters of the example sources we used in this study.  We fixed the angular parameters for each source at the same values: $\left\{ \theta, \ \phi, \ \iota, \ \psi, \ \theta_0 \right\} = \left\{ \frac{\pi}{2}, \ \frac{5\pi}{3}, \ \frac{\pi}{4}, \ \frac{\pi}{3}, \ 1 \right\}$ rad.  For quick reference, we include the coalescence time of each source (computed from equation~\ref{eqn: coalescence time/angle}), a representative value of the Fresnel number (computed from equation~\ref{eqn: Fresnel number}) for that source placed at $R=100$ Mpc (e.g. the approximate distance of the Coma Cluster) with a pulsar at $L=1$ kpc, and the gravitational wavelength of each source (computed from equation~\ref{eqn: gw wavelength}).}
    \label{tab: source parameters reference}
    \begin{tabular}{ c|c|c||c|c|c } 
             \hline
             & \begin{tabular}{c}$\mathcal{M}$ \\ $\left(10^9 \text{ M}_\odot\right)$\end{tabular} & \begin{tabular}{c}$\omega_0$ \\ $\left(\text{nHz}\right)$\end{tabular} & \begin{tabular}{c}$\Delta \tau_c$ \\ $\left(\text{kyr}\right)$\end{tabular} & \begin{tabular}{c} $F$ \\ $\left(L=1\text{ kpc, } \right.$ \\ $\left.R=100 \text{ Mpc}\right)$\end{tabular} & \begin{tabular}{c}$\lambda_\mathrm{gw}$ \\ $\left(\text{pc}\right)$\end{tabular} \\
             \hline
             Source 1 & 10 & 30 & 1.076 & 0.0098 & 1.017 \\
             Source 2 & 0.03 & 20 & $5.085\times10^4$ & 0.0066 & 1.526 \\
             \hline
    \end{tabular}
\end{table}

\begin{table}
\centering
    \caption{This is the pulsar timing array we used in this study, which consists of 40 pulsars from the NANOGrav PTA in~\citet{NG_11yr_data}.  The `Figure-of-merit' (F.O.M.) pulsar distance values were used only in the studies in Section~\ref{subsubsec: Figure-of-Merit}.  Note that the final pulsar has no F.O.M. $L$ value because it was a variable and is indicated in the discussion and figures of that section.  The pulsar data in the $L$, $\theta_p$, and $\phi_p$ columns were taken from version 1.57 of the ATNF catalogue~\citep{ATNF_catalogue}.  For quick reference, we include a representative value of the Fresnel number (computed from equation~\ref{eqn: Fresnel number}) for that pulsar with a source with an orbital frequency of $\omega_0=10$ nHz and a distance of $R=100$ Mpc (e.g. the approximate distance of the Coma Cluster), as well as the monochromatic plane-wave pulsar wrapping cycle distance computed from equation~\ref{eqn:Delta L in monochromatic regimes} for Source 1 (Table~\ref{tab: source parameters reference}).}
    \label{tab: pta reference}
    \begin{tabular}{ l|c|c|c|c|c||c|c } 
            \hline
             & Name & \begin{tabular}{c}F.O.M. $L$ \\ $\left(\text{kpc}\right)$\end{tabular} & \begin{tabular}{c}$L$ \\ $\left(\text{kpc}\right)$\end{tabular} & \begin{tabular}{c}$\theta_p$ \\ $\left(\text{rad}\right)$\end{tabular} & \begin{tabular}{c}$\phi_p$ \\ $\left(\text{rad}\right)$\end{tabular} &  \begin{tabular}{c} $F$ \\ $\left(\omega_0=10 \text{ nHz,}\right.$ \\ $\left.R=100\text{ Mpc} \right)$\end{tabular} &  \begin{tabular}{c} $\Delta L_1$ \\ $\left(\text{pc}\right)$ \\ $\left(\text{Source 1}\right)$ \end{tabular} \\  
             \hline
            1. & J0030+0451 & 0.53 & 0.36 & 1.49 & 0.13 & 0.0004 & 1.65 \\
            2. & J1744-1134 & 0.55 & 0.40 & 1.77 & 4.64 & 0.0005 & 5.39 \\
            3. & J2145-0750 & 0.57 & 0.53 & 1.71 & 5.70 & 0.0009 & 8.89 \\
            4. & J2214+3000 & 0.60 & 0.60 & 1.05 & 5.82 & 0.0012 & 3.68 \\
            5. & J1012+5307 & 0.63 & 0.70 & 0.64 & 2.67 & 0.0016 & 0.68 \\
            6. & J1614-2230 & 0.68 & 0.70 & 1.96 & 4.25 & 0.0016 & 2.08 \\
            7. & J1643-1224 & 0.68 & 0.74 & 1.79 & 4.38 & 0.0018 & 2.82 \\
            8. & J0613-0200 & 0.69 & 0.78 & 1.61 & 1.63 & 0.0020 & 0.54 \\
            9. & J0645+5158 & 0.70 & 0.80 & 0.66 & 1.77 & 0.0021 & 0.64 \\
            10. & J1832-0836 & 0.71 & 0.81 & 1.72 & 4.85 & 0.0021 & 12.13 \\
            11. & J2302+4442 & 0.72 & 0.86 & 0.79 & 6.03 & 0.0024 & 2.03 \\
            12. & J1918-0642 & 0.72 & 0.91 & 1.69 & 5.06 & 0.0027 & 45.35 \\
            13. & J0740+6620 & 0.72 & 0.93 & 0.41 & 2.01 & 0.0028 & 0.73 \\
            14. & J1455-3330 & 0.73 & 1.01 & 2.16 & 3.91 & 0.0033 & 1.27 \\
            15. & J1741+1351 & 0.73 & 1.08 & 1.33 & 4.63 & 0.0038 & 5.04 \\
            16. & J1909-3744 & 0.74 & 1.14 & 2.23 & 5.02 & 0.0043 & 4.46 \\
            17. & J2010-1323 & 0.75 & 1.16 & 1.80 & 5.28 & 0.0044 & 37.55 \\
            18. & J1713+0747 & 0.75 & 1.18 & 1.43 & 4.51 & 0.0046 & 3.92 \\
            19. & J1923+2515 & 0.76 & 1.20 & 1.13 & 5.08 & 0.0047 & 9.55 \\
            20. & B1855+09   & 0.77 & 1.20 & 1.40 & 4.96 & 0.0047 & 19.62 \\
            21. & J1024-0719 & 0.78 & 1.22 & 1.70 & 2.73 & 0.0049 & 0.57 \\
            22. & J0023+0923 & 0.78 & 1.25 & 1.41 & 0.10 & 0.0051 & 1.71 \\
            23. & J2043+1711 & 0.79 & 1.25 & 1.27 & 5.43 & 0.0051 & 16.20 \\
            24. & J1453+1902 & 0.80 & 1.27 & 1.24 & 3.90 & 0.0053 & 1.30 \\
            25. & J1853+1303 & 0.81 & 1.32 & 1.34 & 4.95 & 0.0057 & 15.40 \\
            26. & J1944+0907 & 0.82 & 1.36 & 1.41 & 5.17 & 0.0061 & 67.61 \\
            27. & J2017+0603 & 0.83 & 1.40 & 1.47 & 5.31 & 0.0064 & 130.45 \\
            28. & J2317+1439 & 0.84 & 1.43 & 1.31 & 6.10 & 0.0067 & 2.73 \\
            29. & J1738+0333 & 0.84 & 1.47 & 1.51 & 4.62 & 0.0071 & 5.49 \\
            30. & J1640+2224 & 0.84 & 1.50 & 1.18 & 4.36 & 0.0074 & 2.49 \\
            31. & J1910+1256 & 0.85 & 1.50 & 1.34 & 5.02 & 0.0074 & 20.71 \\
            32. & J0340+4130 & 0.85 & 1.60 & 0.85 & 0.96 & 0.0084 & 0.77 \\
            33. & J2033+1734 & 0.85 & 1.74 & 1.26 & 5.38 & 0.0099 & 17.61 \\
            34. & J1600-3053 & 0.85 & 1.80 & 2.11 & 4.19 & 0.0106 & 1.78 \\
            35. & J2229+2643 & 0.86 & 1.80 & 1.10 & 5.89 & 0.0106 & 3.48 \\
            36. & J1903+0327 & 0.87 & 1.86 & 1.51 & 4.99 & 0.0113 & 31.90 \\
            37. & B1937+21   & 0.88 & 3.50 & 1.19 & 5.15 & 0.0401 & 13.55 \\
            38. & J0931-1902 & 0.88 & 3.72 & 1.90 & 2.49 & 0.0453 & 0.54 \\
            39. & B1953+29   & 0.89 & 6.30 & 1.06 & 5.22 & 0.1300 & 7.96 \\
            40. & J1747-4036 & --- & 7.15 & 2.28 & 4.66 & 0.1675 & 2.80 \\
            \hline
    \end{tabular}
\end{table}


\bsp	
\label{lastpage}
\end{document}


\maketitle
\thispagestyle{empty} 

Here we outline a simple analysis to check that the gravitational wave signal-to-noise ratio (SNR) is not impacted by including other timing effects in the model for our considered sources.  This is important because if for example we considered a low frequency gravitational wave source whose timing residual over our simulated 10 year observation time didn't complete a full wave cycle and appeared parabolic (like a half wave cycle), then our parameter estimation results may be overly optimistic.  This is because a real timing experiment would simultaneously fit a quadratic polynomial and the gravitational wave signal to the timing data (as well as other models which we are not considering for simplicity here).  Therefore we would be less sensitive to a lower frequency source whose gravitational wave timing residual predicted by one of our models from Section 3 resembled a parabola.

In order to help decide what source frequencies we should \textit{not} consider in our Fisher matrix analyses in this paper we first performed the following calculation.  Using following proxy for the SNR of our experiment:
\begin{equation}
    \mathrm{SNR} = \sqrt{\sum_a \left(\frac{\mathrm{Res}_a}{\sigma} \right)^2} ,
\label{eqn: SNR}
\end{equation}
we calculated the SNR of just the gravitational wave timing residuals for every pulsar at every time, using our IIB model (Section 3.4).  Then we recalculated the SNR, but first subtracted off a best-fit quadratic polynomial to the original gravitational wave timing residual:
\begin{equation}
    \mathrm{Res}_\mathrm{polyfit} = \mathrm{Res} - \left( a + bt + ct^2\right) ,
\end{equation}
This polynomial fitting procedure was done for every pulsar in the array, given its timing data, meaning in a PTA of 40 pulsars there were 120 polynomial best-fit parameters.  The result of this is given in Figure~\ref{fig: SNR}, which shows the fractional difference between these two SNR values.  If the fractional difference is low, then we can safely perform our Fisher matrix analysis on that source given that PTA without needing to model the additional parabolic timing residual.  Therefore we considered only sources in this study where the fractional difference in the SNR was small, that is with $\omega_0 > 10$ nHz.
\renewcommand{\thefigure}{1S}
\begin{figure}
    \centering
    \includegraphics[width=0.75\linewidth]{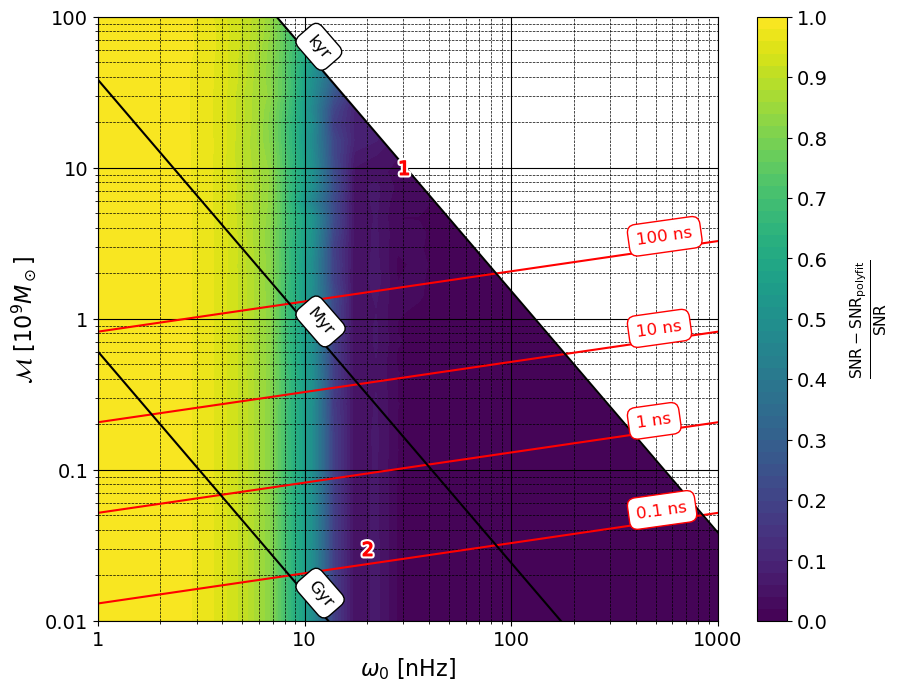}
    \caption{The fractional difference in the gravitational wave timing residual SNR defined in equation~\ref{eqn: SNR}, with vs. without a quadratic polynomial best-fit subtracted from the data, as a function of the source orbital frequency $\omega_0$ and chirp mass $\mathcal{M}$.  The fractional difference in SNR is independent of timing uncertainty $\sigma$, and we also found that it remains effectively unchanged for different values of $R$ (this specific result was computed for $R=100$ Mpc).  For this study we want to consider sources where the fractional difference in SNR is negligible, since we are only modelling gravitational wave effects in our Fisher matrix analyses.  Here we see a low fractional difference of less than $0.1$ at values beyond $\omega_0 > 10$ nHz.  Additionally contours of coalescence time $\Delta \tau_c$ (equation 10) are indicated, with a cut along $\Delta \tau_c = 1$ kyr due to assumption (xiv).  Contours of $A_{E,\mathrm{res}}\left(\frac{R}{100 \ \mathrm{Mpc}}\right)$ are also indicated in red, which serve as a useful proxy for how strong the timing residual signal is (especially when compared to the timing noise $\sigma$).  The fiducial sources from Table B1 are also indicated by their number 1 and 2 here, and have fractional SNR differences here of approximately $0.03$ and $0.06$, respectively.  This was computed using the IIB model (Section 3.4).}
    \label{fig: SNR}
\end{figure}

Further restrictions on the searchable region of parameter space come from consideration of the coalescence time (our fundamental assumption (xiv)), as well as the strength of the timing residual, which as a proxy can be estimated by the parameter $A_{E,\mathrm{res}}$ (equation~38).  These are also indicated in Figure~\ref{fig: SNR}.

\newpage

\renewcommand{\thefigure}{2S}
\begin{figure}
    \centering
    \includegraphics[width=1\linewidth]{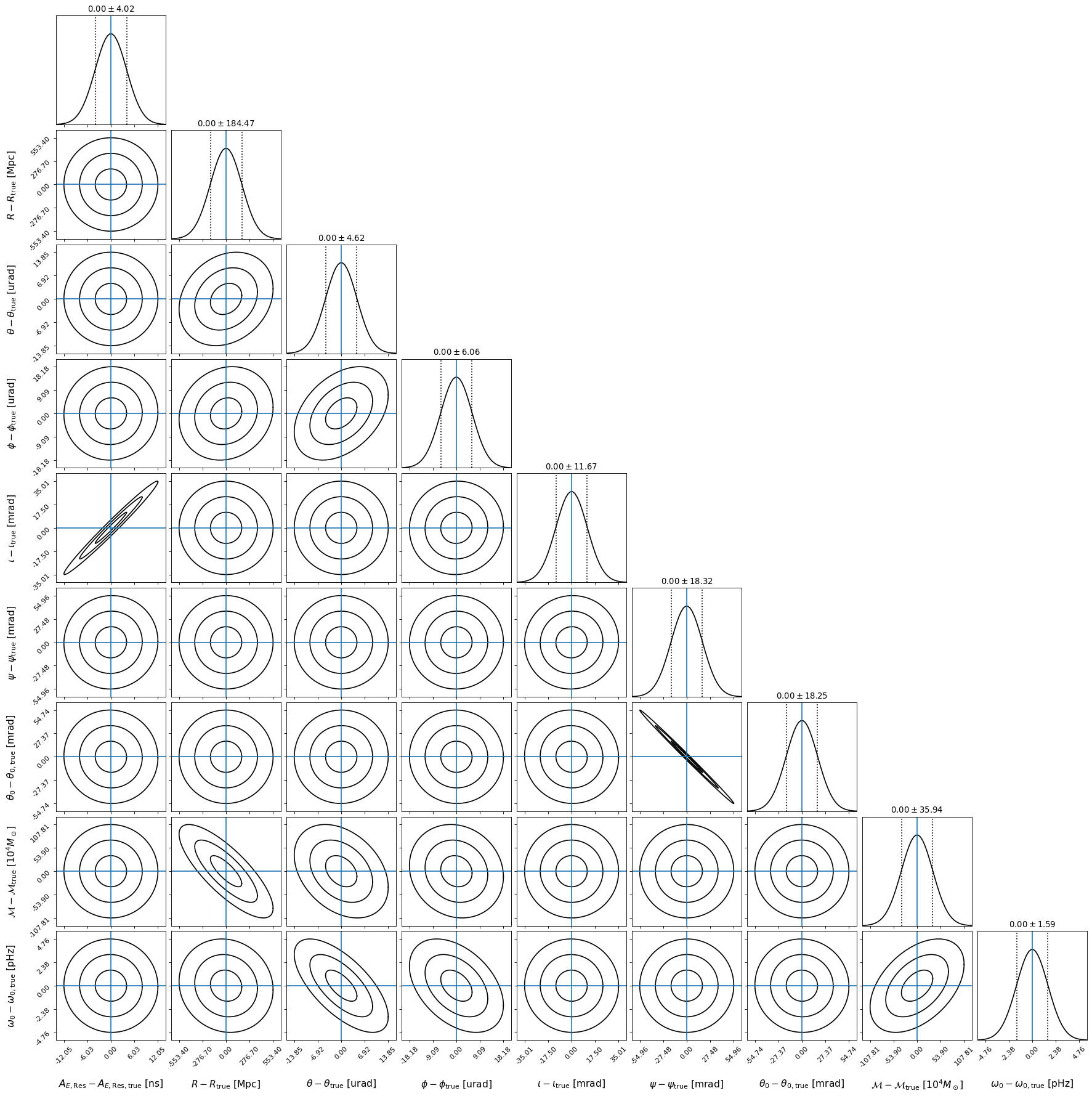}
    \caption{Fisher matrix measurements of the parameters for Source 1, which has been placed at $R=500$ Mpc (therefore $A_{E,\mathrm{res}}=414$ ns).  Here the PTA has a scale factor of $7$ and all pulsars have been given $\sigma_L = 1 \ \mathrm{pc}\ \sim \lambda_\mathrm{gw}$, with timing uncertainty $\sigma = 100$ ns.  This simulation and experimental set-up corresponds to the red square indicated in Figure 5.  The Fisher matrix analysis predicts that this source can be recovered with:
    $\left\{ \mathrm{CV}_{A_{E,\mathrm{res}}}, \ \mathrm{CV}_R, \ \mathrm{CV}_\theta, \ \mathrm{CV}_\phi, \ \mathrm{CV}_\iota, \ \mathrm{CV}_\psi, \ \mathrm{CV}_{\theta_0}, \ \mathrm{CV}_\mathcal{M}, \ \mathrm{CV}_{\omega_0} \right\} = \left\{ 0.0097, \ 0.367, \ 3\times10^{-6}, \ 10^{-6}, \ 0.0148, \ 0.0175, \ 0.0182, \ 3.6\times10^{-5}, \ 5.3\times10^{-5} \right\}$.  These results were computed using the IIB model (Section 3.4).}
    \label{fig: s1 example triangle}
\end{figure}